\documentclass[sigconf]{acmart}
\AtBeginDocument{%
  }

\copyrightyear{2026}
\acmYear{2026}
\setcopyright{cc}
\setcctype{by}
\acmConference[CCS '26]{Proceedings of the 2026 ACM SIGSAC Conference on Computer and Communications Security}{November 15--19, 2026}{The Hague, Netherlands}
\acmBooktitle{Proceedings of the 2026 ACM SIGSAC Conference on Computer and Communications Security (CCS '26), November 15--19, 2026, The Hague, Netherlands}
\acmDOI{10.1145/3830454.3832715}
\acmISBN{979-8-4007-2871-6/2026/11}

\usepackage{enumerate}
\usepackage{enumitem}
\usepackage{graphicx}
\usepackage{epstopdf}
\usepackage{epsfig}
\usepackage{etoolbox}
\usepackage{tikz}
\usepackage{amsmath}
\usepackage{stmaryrd}
\usepackage{mathtools}
\usepackage{hhline}
\usepackage{setspace}
\usepackage{booktabs}
\usepackage{multicol}
\usepackage{multirow}
\usepackage{xcolor,colortbl}
\usepackage{array}
\usepackage{xspace}
\usepackage{pifont}
\usepackage{graphicx}
\usepackage{subcaption}
\usepackage[most]{tcolorbox}
\usepackage[htt]{hyphenat}

\usepackage[flushleft]{threeparttable}
\usepackage[multiple]{footmisc}
\usepackage{adjustbox}
\usepackage{diagbox}
\usetikzlibrary{arrows}
\usepackage{wasysym}
\usepackage{longtable}

\usepackage{listings}
\usepackage{color}
\usepackage{enumitem}

\usepackage{fontawesome5}
\usepackage{cleveref}

\definecolor{dkgreen}{rgb}{0,0.6,0}
\definecolor{gray}{rgb}{0.5,0.5,0.5}
\definecolor{mauve}{rgb}{0.58,0,0.82}
\definecolor{Gray}{gray}{0.9}
\definecolor{persiangreen}{rgb}{0.0, 0.65, 0.58}
\definecolor{viridian}{rgb}{0.25, 0.51, 0.43}
\definecolor{ruddy}{rgb}{1.0, 0.0, 0.16}

\newcommand{\mac}{macOS\xspace}
\newcommand{\system}{NutriScan\xspace}

\newcommand{\manifesttotal}{588\xspace}

\newcommand{\entitlements}{828\xspace}
\newcommand{\nls}{Nutrition Labels\xspace}
\newcommand{\nl}{Nutrition Label\xspace}

\DeclareRobustCommand*\circledchar[2]{
  \tikz[baseline=(c.base)]
    \node[circle, white, font=\sffamily\bfseries, fill={#1},
    draw={#1}, inner sep=0pt, minimum size=1em](c){#2};
}

\newcommand*\circlered[1]{\circledchar{red}{#1}}

\definecolor{borderblue}{HTML}{006EAF}
\definecolor{fillblue}{HTML}{1BA1E2}
\DeclareRobustCommand*\circledbluechar[2]{%
  \leavevmode
  \tikz[baseline=(c.base)]{%
    \node[
      circle,
      white,
      font=\sffamily,
      fill={#1},
      draw={borderblue},
      inner sep=0pt,
      minimum size=1em
    ] (c) {\footnotesize#2};
  }%
  \unskip\nobreak
}
\newcommand*\circleblue[1]{\circledbluechar{fillblue}{#1}\ignorespaces}

\usepackage{pgfplots}
\pgfplotsset{compat=1.16}

\definecolor{lightgray}{rgb}{0.95, 0.95, 0.95}
\definecolor{darkgray}{rgb}{0.4, 0.4, 0.4}
\definecolor{purple}{rgb}{0.65, 0.12, 0.82}
\definecolor{editorOcher}{rgb}{1, 0.5, 0} 
\definecolor{editorGreen}{rgb}{0, 0.5, 0} 
\definecolor{orange}{rgb}{1,0.45,0.13}		
\definecolor{olive}{rgb}{0.17,0.59,0.20}
\definecolor{brown}{rgb}{0.69,0.31,0.31}
\definecolor{purple}{rgb}{0.38,0.18,0.81}
\definecolor{lightblue}{rgb}{0.1,0.57,0.7}
\definecolor{lightred}{rgb}{1,0.4,0.5}

\definecolor{darkpurple}{rgb}{0.502, 0.26, 0.747}
\definecolor{background}{HTML}{EEEEEE}
\definecolor{commentgreen}{RGB}{2,112,10}
\colorlet{numb}{magenta!60!black}
\definecolor{eminence}{RGB}{108,48,130}

\usepackage{xurl}

\usepackage{hyperref}
\hypersetup{
  hidelinks,
	linkcolor = {darkpurple},
  breaklinks = true
}
\usepackage{upquote}

\definecolor{eclipseStrings}{RGB}{42,0.0,255}
\definecolor{eclipseKeywords}{RGB}{127,0,85}
\colorlet{numb}{magenta!60!black}
\lstdefinelanguage{json}{
  keywordstyle=\color{blue}\bfseries,
  keywords={permissions, content\_scripts,manifest\_version,background,browser\_action},
  keywordstyle=\color{lightblue},
  identifierstyle=\color{black},
  sensitive=false,
  comment=[l]{//},
  morecomment=[s]{/*}{*/},
  commentstyle=\color{purple}\ttfamily,
  stringstyle=\color{red}\ttfamily,
  frame=single,
    literate=
     *{0}{{{\color{numb}0}}}{1}
      {1}{{{\color{numb}1}}}{1}
      {2}{{{\color{numb}2}}}{1}
      {3}{{{\color{numb}3}}}{1}
      {4}{{{\color{numb}4}}}{1}
      {5}{{{\color{numb}5}}}{1}
      {6}{{{\color{numb}6}}}{1}
      {7}{{{\color{numb}7}}}{1}
      {8}{{{\color{numb}8}}}{1}
      {9}{{{\color{numb}9}}}{1}
      {\{}{{{\color{delim}{\{}}}}{1}
      {\}}{{{\color{delim}{\}}}}}{1}
      {[}{{{\color{delim}{[}}}}{1}
      {]}{{{\color{delim}{]}}}}{1},
}

\definecolor{delim}{RGB}{20,105,176}
\definecolor{numb}{RGB}{106, 109, 32}
\definecolor{string}{rgb}{0.64,0.08,0.08}

\lstdefinelanguage{Swift}{
  keywords={typeof, new, true, false, catch, function, func, async, return, null, catch, switch, var, if, in, while, do, else, case, break, AVCaptureSession, AVCaptureDeviceInput, AVCaptureDevice, video, authorized, notDetermined},
  keywordstyle=\color{blue}\bfseries,
  ndkeywords={class, export, boolean, throw, implements, import, this},
  ndkeywordstyle=\color{darkgray}\bfseries,
  identifierstyle=\color{black},
  sensitive=false,
  comment=[l]{//},
  morecomment=[s]{/*}{*/},
  commentstyle=\color{purple}\ttfamily,
  frame=single,
  stringstyle=\color{red}\ttfamily,
  morestring=[b]',
  morestring=[b]"
}

\lstdefinelanguage{json}{
  keywordstyle=\color{blue}\bfseries,
  keywords={Camera, Data,Type, Purpose, Handling, User, Permission, API, Access, true, false, notdetermined},
  keywordstyle=\color{lightblue},
  identifierstyle=\color{black},
  sensitive=false,
  comment=[l]{//},
  morecomment=[s]{/*}{*/},
  commentstyle=\color{purple}\ttfamily,
  stringstyle=\color{red}\ttfamily,
  frame=single,
    literate=
     *{0}{{{\color{numb}0}}}{1}
      {1}{{{\color{numb}1}}}{1}
      {2}{{{\color{numb}2}}}{1}
      {3}{{{\color{numb}3}}}{1}
      {4}{{{\color{numb}4}}}{1}
      {5}{{{\color{numb}5}}}{1}
      {6}{{{\color{numb}6}}}{1}
      {7}{{{\color{numb}7}}}{1}
      {8}{{{\color{numb}8}}}{1}
      {9}{{{\color{numb}9}}}{1}
      {\{}{{{\color{delim}{\{}}}}{1}
      {\}}{{{\color{delim}{\}}}}}{1}
      {[}{{{\color{delim}{[}}}}{1}
      {]}{{{\color{delim}{]}}}}{1},
}

\begin{document}

\title[Exploring Privacy Leakage and Data Disclosure Violations in the MacOS Application Ecosystem]{Exploring Privacy Leakage and Data Disclosure Violations in the MacOS Application Ecosystem}
\author{Jyotirmay Chauhan}
\affiliation{%
\institution{University of Illinois Chicago}
\city{Chicago}
\state{Illinois}
\country{USA}}
\email{jchauh3@uic.edu}

\author{Kostas Solomos}
\affiliation{%
\institution{Brandeis University}
\city{Waltham}
\state{Massachusetts}
\country{USA}}
\email{solomos@brandeis.edu}

\author{Mir Masood Ali}
\affiliation{%
\institution{University of Illinois Chicago}
\city{Chicago}
\state{Illinois}
\country{USA}}
\email{mali92@uic.edu}

\author{Jason Polakis}
\affiliation{%
\institution{University of Illinois Chicago}
\city{Chicago}
\state{Illinois}
\country{USA}}
\email{polakis@uic.edu}

\begin{abstract}

The systematic and excessive data collection practices of tech companies have
rendered online privacy both a necessity and a sought-after commodity.
However, while the privacy risks of the web, mobile, and IoT ecosystems
have been extensively examined, desktop environments have been largely
overlooked.
As desktop apps continue to be widely used,
they remain a critical yet
understudied dimension of user privacy.
In this paper, we address this gap by presenting the first, to our knowledge,
comprehensive study of the mechanisms designed to regulate and disclose
data collection and sharing practices in the macOS ecosystem. We adopt an
app-development-centric view, and shed light on the interactions between
the various macOS mechanisms that mediate apps' data access.
Driven by our findings, we develop \system, an analysis framework that
incorporates both static and dynamic analysis techniques to create a
consolidated view of macOS apps' data practices and disclosures.
We use our system to dynamically analyze
1K macOS apps,
and find that 85\% of them access user-data APIs without
disclosing it. 49.7\% also exfiltrate data to advertising
entities and hosting providers, 12.5\% of which do so without a corresponding disclosure.
We find that desktop apps are being leveraged by online
trackers to enrich user profiles and device fingerprints, thus 
shedding new light on the true scope of the
online tracking ecosystem.
Our analysis reveals how the macOS app ecosystem is
comprised of disjoint mechanisms with divergent data abstractions,
thus increasing complexity for developers while also facilitating undisclosed
privacy-invasive practices. Accordingly, we propose a series of mitigations
that aim to both streamline the data disclosure 
process for developers and improve Apple's app vetting process.

\end{abstract}

\maketitle

\section{Introduction}
\label{sec:intro}

The proliferation of smartphones and
Internet-connected devices has resulted in users constantly
interacting with devices that generate, store, and transfer a plethora of
private data. This abundance of data has given rise to surveillance
capitalism~\cite{zuboff2023age}, a complex ecosystem powered by pervasive
collection and opaque sharing practices, where user data is brokered among
corporations~\cite{bashir-pets-2018,197268}. 
The collected data enables the inference of sensitive and
\emph{protected}~\cite{protected-characteristics} user characteristics and
behaviors~\cite{pmlr-v81-speicher18a,castelluccia2012betrayed}, highly
effective advertising~\cite{10.1145/2187836.2187852,10.1145/1526709.1526745}
and misinformation campaigns~\cite{dawson2021microtargeting}.

Users have since grown increasingly concerned about
privacy~\cite{wijesekera2017feasibility,cummings2021need}, calling for better
technical and legislative measures~\cite{203840}. Consequently, there have been
concerted efforts by browser and operating system (OS) vendors for the
deployment of more effective anti-tracking and access control mechanisms,
respectively~\cite{privacysandbox,firefoxETP,applekit}. At the same time, OS
vendors have focused on deploying mechanisms that enable the disclosure
of data practices in a more user-friendly manner, as opposed to the confusing
\emph{legalese} found in privacy policies. More specifically, Apple's Privacy
Nutrition Labels and Google's Data Safety Sections provide the means for app
developers to inform users about the data they collect, whether they share it
with other parties, and how it is used~\cite{manifest_announcement,datasafety}.

While the web, mobile, and IoT ecosystems have been extensively studied in
regard to potential privacy leakage and
violations~\cite{wang2025privacyguard,ullah2023privacy,
    lalaine,arkalakis2024abandon,kollnig2021fait,linden2020privacy,mobileWebAPIAttacks2019,acar2013fpdetective,
    das2016tracking,10.1145/2906388.2906392,ren2018longitudinal,anselmi2024watching,iqbal2023tracking,ali2023problematic,zeng2021makes},
the desktop ecosystem has been largely overlooked even though 61.5\% of online
adults accessed the internet via desktop devices in Q3 2024, including 72.7\%
in the United States~\cite{desktop-use}.
This oversight could be attributed to desktop apps not usually displaying ads
and hence not typically being associated with privacy leakage. Nonetheless,
it is currently unclear what role desktop apps play in the tracking and
advertising landscape. 
Desktops fundamentally allow broader default access to 
user data, and macOS introduces a unique security and permission architecture that differs from iOS. These differences collectively motivate our study. Specifically,
this includes (i)~understanding what access control and
disclosure mechanisms exist in a modern desktop OS, (ii)~investigating whether
desktop apps employ tracking and fingerprinting techniques, and (iii)~verifying
apps' data practice disclosures. 

In this paper, we fill this gap by presenting the first, to the best of our
knowledge, exploration of privacy leakage in desktop applications. We focus
specifically on the macOS ecosystem, due to users perceiving it as a more
private operating system compared to Windows~\cite{topcuoglumacos}, the
presence of multiple OS-level security and privacy
mechanisms~\cite{apple-security}, and Apple's recent mandatory requirement for
apps to release Privacy Nutrition Labels~\cite{apple_nutrition_labels}. To that
end, we develop \system, a novel framework that leverages both static and
dynamic analysis techniques to consolidate information across all relevant application
components. This includes
analyzing bundle components that define installation and run-time permissions
for data access and protected system resources, and dynamically instrumenting
native APIs to identify what data and capabilities are made available to the
application. We also augment our analysis with network traffic to identify what
protected data is transferred off the device. Finally, our system generates
\nls~that accurately capture app behavior, and flags inconsistencies in
official \nls.

Overall, our experiments reveal a troubling pattern of overly intrusive data
collection practices alongside incomplete and incorrect data disclosures. Our
data suggests that privacy screening procedures currently in place (if any) are
insufficient, as developers often fail to disclose what data is being collected
and shared by both first-party and third-party components, as reflected by the
inconsistencies captured by \system. Specifically, we find that only $\sim15\%$
of the tested apps correctly disclosed all data labels associated with the
sensitive data they accessed. Crucially, developers did not account for the
data collected by embedded third-party code in over 80\% of the apps with
manifest files. More importantly, we found sensitive user data, and immutable
user and device identifiers being shared with advertising networks, including
Google and Facebook.
As such, despite Apple’s recent marketing efforts showcasing privacy as a major
selling point, we find that even within their heavily guarded desktop
environment, data protection and disclosure mechanisms are currently
insufficient, necessitating a redesign of their mechanisms and a more stringent
vetting process.

In summary, our major research contributions are:
\begin{itemize}[nosep,noitemsep]

    \item We explore the mechanisms that govern data collection and disclosure practices
        in the macOS ecosystem.

    \item We develop \system, a novel dynamic analysis framework that comprehensively
          evaluates macOS applications for incomplete or incorrect data disclosures. We
          have open sourced our system to enable developers to accurately report their
          applications' data practices.~\cite{coderepo}.

    \item We conduct the first study of tracking and fingerprinting in desktop apps, and
          uncover the prevalence of inaccuracies and omissions in macOS' Privacy
          Nutrition Labels. 

    \item We propose a unified data disclosure and verification process for
          mitigating the privacy violations that stem from the complex and disjoint
          nature of macOS mechanisms.

\end{itemize}

\begin{figure}[t!]
  \centering
  \includegraphics[width=1.0\columnwidth]{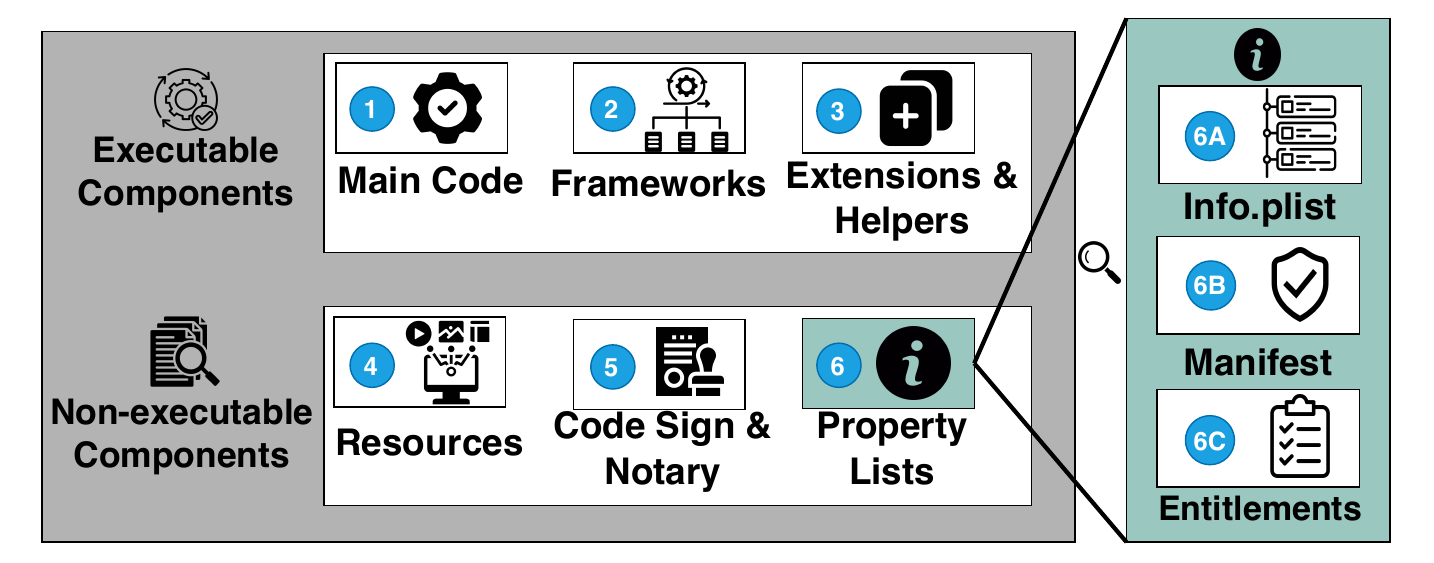}
  \caption{Application bundle structure and components.}
  \label{fig:app_bundle}
\end{figure}

\section{Background and Threat Model}
\label{sec:background}

Here we provide background information on macOS and its app ecosystem, which is
pertinent to our work.  Figure~\ref{fig:app_bundle} illustrates a macOS
application bundle and its components, resources, and associated metadata.

\subsection{App Structure and Lifecycle.}
macOS applications are built, distributed, and executed within Apple's
ecosystem that integrates dedicated development tools, distribution
infrastructure, and runtime enforcement mechanisms. Apple provides developers
with \texttt{Xcode}, an Integrated Development Environment (IDE), which
includes interface design, coding, and resource management, as well as integral
tools for code signing and configuration~\cite{apple_xcode}. App developers
produce their own first-party code~\textbf{\circleblue{1}} and also include
third-party SDKs~\textbf{\circleblue{2}} during the development process. When
an application is production-ready, Xcode packages the contents into a
structured \texttt{bundle}, which developers can submit to the App
Store~\cite{apple_bundle}.

A key component of the bundle is the Information Property List
(\texttt{Info.plist})~\textbf{\circleblue{6A}}, which specifies app-related
metadata, such as the bundle ID, version number, supported platforms, URL
schemes, and document types~\cite{apple_plist}. Developers are also required to
define security and privacy-relevant declarations to access protected system
resources (e.g.,~camera, location). \texttt{Entitlements} define the
capabilities that the app requests that require elevated
privileges~\cite{apple_entitlements}~\textbf{\circleblue{6C}}. System services
use them to grant the app access to restricted APIs or privileged subsystems,
such as iCloud storage, push notifications, or hardware devices. Another key
component is the \texttt{Privacy
  Manifest}~(\texttt{.xcprivacy})~\textbf{\circleblue{6B}}, a structured file
declaring data practices for the app and embedded SDKs, using structured key
values. Developers specify whether the app or an embedded SDK engages in
tracking, and enumerate the specific data types collected (e.g.,~device
identifiers, coarse location, phone number). Each data item is individually
labeled to indicate whether it is ``linked'' to the user and annotated with the
purposes for which it is collected, such as analytics, app functionality, or
advertising~\cite{apple_privacy_manifest}. \texttt{Xcode} aggregates these
manifests into a single report at submission
time~\cite{apple_privacy_manifest}.

Additionally, the \mac~security model verifies the app at three distinct stages: (i)~during
the build via Xcode, (ii)~during distribution through the App Store Connect
submission and evaluation system, and (iii)~at runtime through signature and
integrity checks. Each step corresponds to distinct cross-stage verification
and enforcement checks applied to bundle components, including metadata 
that define the app's identity and capabilities, declarations of access to system resources, and disclosures of data collection practices.

\subsubsection{\emph{\faWrench~Build-time Verification.}} When developers build an app, \texttt{Xcode}
assembles the source code, static resources, and configuration files into the
bundle where different Apple mechanisms verify their validity and correctness.
Specifically, \texttt{Xcode} validates \texttt{Info.plist} entries structurally and
semantically~\textbf{\circleblue{6A}}, ensuring that declared access to protected resources
(e.g.,~camera, location) is present as required for runtime access.
During development, a developer may insert permission entries to the
\texttt{Info.plist}
file and \texttt{Xcode} automatically signs the binary with the necessary entitlements~\textbf{\circleblue{6C}}.
It also includes a \texttt{CodeResources} file containing the file hashes in the bundle~\textbf{\circleblue{4}}
and signs it to ensure the app's integrity.  When the developer is
ready to ship the app, they specify the team identifier, which allows \texttt{Xcode} to
sign the bundle with the Apple Development certificate~\textbf{\circleblue{5}}, allowing the app to run on
other systems.

\subsubsection{ \emph{\faUpload~Submission \& Compliance.}} Upon submission to the App Store,
Apple performs a combination of automated and manual reviews to enforce
platform-specific policies~\cite{app_review}. This process aims to verify
code-signing integrity, validate entitlements, and ensure compliance with
sandboxing requirements. Developers are also required to submit
the app's  Nutrition Labels. The labels enumerate the categories of data
collected (e.g., contact information, location, device identifiers), specify
whether each category is associated with the user's identity, and if they are
used for tracking~\cite{apple_data_tracking_linking}. Labels are
structured into three groups: (i)~\texttt{Data Linked to You} (information associated
with a personal identity), (ii)~\texttt{Data Not Linked to You} (information purported
to be anonymized), and (iii)~\texttt{Data Used to Track You} (information correlated
across apps or services). An example can be found in Figure~\ref{fig:capcut} (Appendix~\ref{appendix:nutrition}).

\subsubsection{  \emph{\faRocket~Runtime Validation.}} When the app first launches, the system
validates its origin and integrity. This
includes code-signing checks and team identifier checks.
The code is also checked for malware signatures by
\texttt{Gatekeeper} and \texttt{XProtect}, Apple's dedicated verification
systems~\cite{gatekeeper}. At runtime, access to privacy-sensitive
resources (e.g.,~camera, microphone) is verified by the Transparency, Consent,
and Control (\texttt{TCC}) framework~\cite{apple_tcc} which combines checks
with user prompts, verifying property list files and the system's consent
database before granting access.

\subsubsection{\emph{\faEye~Data Privacy Communication.}} Apple enforces a disclosure system
that separates reviewer-facing and user-facing privacy information.
\texttt{Privacy Manifests} provide machine-readable declarations used during the App
Store review~\textbf{\circleblue{6B}}, while \texttt{Privacy Nutrition Labels} present a simplified,
human-readable summary for users. These mechanisms operate independently,
and Apple \emph{does not} perform consistency checks between them,
which can result in conflicting or inaccurate disclosures. On iOS, developers
must specify predefined ``required reason APIs'' when accessing
fingerprinting-related interfaces~\cite{ios_manifest}. This requirement does
not extend to \mac~\cite{apple_privacy_manifest}. Instead, within the macOS
ecosystem, App Store disclosures coexist with broader system-level enforcement mechanisms, resulting  in a distinct enforcement
environment.

\subsection{macOS vs iOS.}
Mobile operating systems have diverged considerably from traditional desktop
operating systems, especially with regard to application isolation and data
access control~\cite{roesner2012user,spensky2016sok,botha2009desktop}.
Here we provide additional details about macOS and highlight some of the key
differences with iOS.

\emph{\faCodepen~App Sandbox.} macOS implements a sandboxing system for controlling
applications’ behavior and capabilities within the system.  Sandboxed apps are
restricted from accessing files owned by other apps or the user, unless
explicitly permitted (e.g., through system file-selection dialogs).  The
sandbox is enforced by the kernel using the Mandatory Access Control Framework
(MACF), which checks policy violations at runtime.
Since executing apps distributed outside the App Store is common practice in
macOS, it provides more granularity over specific rules and constraints, which
can be defined inside \texttt{.sb} files.  Nonetheless, for convenience and
standardization, Apple provides a default Sandboxing (\texttt{App Sandbox}),
applied via the \texttt{com.apple.security.app-sandbox} entitlement, which is
mandatory for all apps downloaded via the App Store. Apps distributed through
other sources are not required to import this entitlement.
Furthermore, each sandboxed app contains a filesystem hosted at
\texttt{~/Library/Containers} which mirrors a default user's home directory
structure (e.g., \texttt{Downloads}, \texttt{Desktop}). These directories are
empty by default, and the app must request permission via the TCC to obtain
access.  Similarly, access to networking capabilities and Photos is also
restricted by a combination of entitlements and TCC-prompts.

The macOS and iOS sandboxes differ significantly in several key aspects. While iOS enforces a stricter, more contained sandbox model, macOS adopts a more flexible approach.
Apps on iOS execute in strictly isolated containers, while apps on macOS can access standard user location and user-selected files.
Developers on macOS also have access to several
entitlements that provide exceptions to the App
Sandbox~\cite{sandbox-exceptions}. 
Notably, there is no equivalent mechanism on iOS.
These entitlements provide access to the main file system (\texttt{apple.security.temporary-exception.files.home-relative-path}) 
and the IPC system (
\texttt{security.temporary-exception.mach-register.global-name}).
Apps can use exception entitlements to directly control the
system via Apple Events, a macOS-specific automation feature. 
Finally, app sandboxing is not mandatory on macOS, while it is enforced for  the App Store applications on iOS.

\begin{figure*}[!ht]
  \centering
  \includegraphics[width=0.8\textwidth,keepaspectratio]{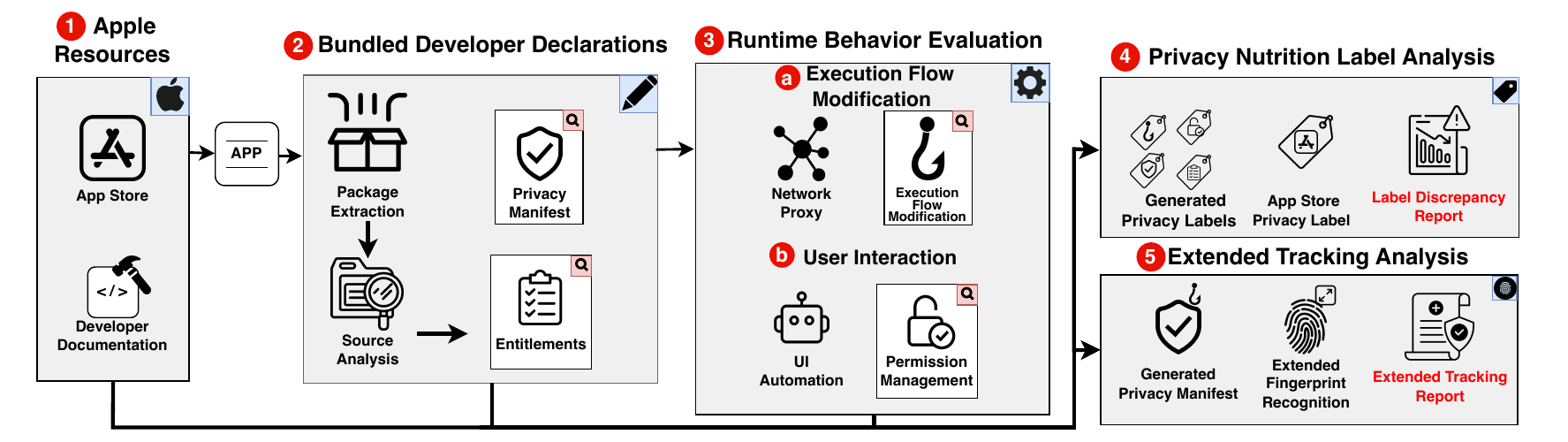}
  \caption{An overview of \system's components and workflow for evaluating \mac~apps.}
  \label{fig:framework}
\end{figure*}

\emph{\faList~System settings storage.} UserDefaults (also referred to as NSUserDefaults)
is a key-value storage mechanism for storing application preferences across restarts and OS reboots~\cite{nsuserdefaults}. It is an app-specific storage designed to record configuration data, such as UI preferences or language settings. 
For sandboxed apps, the UserDefaults store is saved under the
\texttt{Preferences} subdirectory inside the app's container. 
On both macOS and iOS, the UserDefaults file is prepopulated with specific system-derived default values (e.g., locale-related settings) required for the intended app functionality.
Notably, Apple acknowledges that the contents of UserDefaults may reveal
environment-specific information that can be leveraged for fingerprinting~\cite{nsuserdefaults}.

\emph{Summary.} While macOS and iOS share several architectural and
implementation features at both the system level (e.g., kernel, XPC Mach
messaging, executable format) and the application level (e.g., app bundles,
entitlements, TCC permissions), they are \emph{fundamentally} designed around
different security models. iOS follows a highly restrictive, closed ecosystem
model, in which Apple enforces mandatory code signing for all executable code
and discourages application distribution outside official channels. Permissions
on iOS are tightly constrained, and apps cannot access arbitrary filesystem
locations outside their containers without explicit user authorization.
Crucially, iOS does not expose a multi-user account model and  each app
operates under a constrained single default user context, reflecting its design
as a single-user system. 

In contrast, macOS is designed to support more flexible and customizable
workflows.  It  supports  multiple user accounts and permits application
distribution outside the App Store, allowing users to download and execute
unsigned or non–App Store–signed code.  While macOS employs mechanisms such as
Gatekeeper and notarization to mitigate malware risk, these protections are not
uniformly enforced across all execution paths, reflecting support for a broader
set of development and user workflows.  As a result, macOS must address  both
privacy and system security threats.  Apps operate with broader privileges than
on iOS, including access to user-selected files, system automation interfaces,
and native APIs that are not available on the mobile platform.  While prior work
has primarily focused on privacy label compliance in the iOS ecosystem, macOS
applications operate under a more permissive security model that both extends
previously studied attack surfaces and enables additional privacy-invasive
risks.

\subsection{Motivation and Threat Model.}
It is unclear what role, if any, desktop apps play within the broader online
tracking ecosystem. Moreover, several of their characteristics are of
particular interest when considering potential privacy implications. First,
they provide direct access to native APIs that offer a plethora of
device-specific information and immutable identifiers, which can be used by
trackers to augment user and browser-fingerprint profiles.
Second, users typically log into desktop apps, thereby providing the identifier
(i.e., email) necessary for linking together the user's various
devices~\cite{neufeld2017cross,cartology}. The risk of cross-device tracking is
evident in FTC reports and prior studies that explored how companies employ
cross-device tracking techniques to correlate different devices that belong to
the same user~\cite{federal2017cross,zimmeck2017privacy,brookman2017cross}.
Third, desktop apps can access other sensitive user data (e.g.,~address books,
photo libraries) which can enable diverse privacy-invasive inferences. We provide a
detailed breakdown of native API capabilities compared to browsers in
Appendix~\ref{appendix:capabilities}, highlighting how these capabilities can
be leveraged for extending fingerprinting beyond the browser boundaries.

The privacy implications of these characteristics become even more pronounced
when considering our analysis of Apple's data access mechanisms. More
specifically, while App Store reviewers verify the contents of the consolidated
app privacy report generated from individual privacy manifest files, they do
not use this information to verify the accuracy of the Privacy Nutrition Labels
shown to users. This is due to the privacy manifest itself lacking a
requirement for explicitly declaring reasons for accessing sensitive APIs
(which can be used to fingerprint users, as we discuss in~\S\ref{subsec:fp}).
While no prior work has evaluated the accuracy and completeness of \mac~apps'
Privacy Nutrition Labels, studies have shown that labels for mobile apps are
inaccurate~\cite{arkalakis2024abandon,lalaine}, further motivating our
research.

In this study, we are interested in privacy-invasive practices by first- and
third-party code. We focus on a wide range of user and device-specific data
that can be obtained by apps through native APIs, and misused for various
scenarios. Indicatively, we highlight the following scenarios that capture
different privacy-invasive behaviors: (i)~enriching existing user ad profiles,
(ii)~correlating different devices operated by the same user, and
(iii)~deanonymizing an active private browsing session -- e.g., code by
\texttt{Tracker} is embedded in the active \texttt{FOOBAR} desktop app, while
the user is also browsing \texttt{www.example.com} (which includes an
\texttt{iframe} from \texttt{Tracker}) over incognito
mode~\cite{10.1145/3178876.3186088} in their browser, allowing Tracker to
deanonymize the private browsing session. Overall, we are interested in
capturing privacy-invasive practices that \emph{are not} accurately disclosed (or covered
by) apps' Privacy Nutrition Labels.

\section{\system: Design and Implementation}
\label{sec:implementation}

In this section, we detail the design and implementation of our framework,
\system. Figure~\ref{fig:framework} shows an overview of the framework's
components and workflow, which we detail next.

\noindent \textbf{\circlered{1} Apple Resources.} Apple lists and distributes
bundled \mac~apps via its App Store (\faAppStore), which also hosts apps for
other platforms, including those aimed at iOS and iPadOS. Apple requires that
developers submitting to the App Store ensure that the Nutrition Label
and Privacy Manifest, among other declarations, reflect the data collection
practices relevant to the submitted version of the app. To ensure the validity
of our evaluation, we access app bundles directly from the App Store,
instead of using third-party repositories. We also collect the
corresponding user-facing Privacy Nutrition Labels that the developer has
declared on the App Store.

\noindent \textbf{\circlered{2} Bundled Developer Declarations.} Apps
downloaded from the App Store are bundled and structured in a standard format
(see~\S\ref{sec:background}). MacOS apps declare metadata relevant to the system
by listing key-value pairs in
\texttt{Info.plist}, from which we extract the app's
bundle identifier, version, and other metadata.

\textbf{\faSearch~\emph{Privacy Manifest.}} Apple holds developers accountable
for \emph{all} of the code included in their app, including any third-party
SDKs they include~\cite{sdk_requirements}.
However, the size and complex nature of these third-party codebases make it
difficult for developers to independently ensure accuracy and completeness.
Apple, therefore, mandates that third-party SDKs disclose their own data collection
practices via a key-value list in a \texttt{privacy manifest} file, named
\texttt{PrivacyInfo.xcprivacy}. When bundling the app, XCode~($>=$v15) includes
manifests from all third-party SDKs in the output. We extract and parse all
available privacy manifest files, which we then map to the corresponding
entries in the Nutrition Labels.

To compare manifests with Nutrition Labels, we create a union of all the
manifest files present inside the app. This final file contains a superset of
all \texttt{NSPrivacyCollectedDataType} values which we then compare against the
application's Nutrition Labels. We consider the presence of a manifest-declared data type
(e.g., \texttt{NSPrivacyCollectedDataTypeName}) coupled with the absence of the
corresponding data item (e.g., \texttt{Name}) from the Nutrition Labels as
a discrepancy/violation. We note that we only flag a discrepancy
when a Privacy Manifest item is found to be missing from the Nutrition Label,
but not the other way around.

\textbf{\faSearch~\emph{Entitlements.}} While building the app, developers can
add \texttt{Capabilities} to help the app access certain system resources 
(e.g.,~autofill, access to background tasks). 
The contents of these entitlements help us evaluate a set of sensitive
resources that the app accesses, a subset of which maps to entries in the
Nutrition Labels. We parse Apple's Documentation\cite{apple_entitlements} and
third-party sources~\cite{entitlement_3party} to assemble a definitive list of
entitlements. Since several entitlements remain undocumented by Apple, we
cannot guarantee completeness in our list of entitlements; however, our final
list of entitlements was sufficiently complete to allow us to parse \emph{all}
of the entitlements encountered when analyzing the apps in our dataset (see \S
\ref{sec:evaluation}).  We also map entitlements to Nutrition Labels' data
items such as name, physical address, coarse location etc. This mapping was
done independently by two researchers, who provided the same mapping for over
90\% of the cases.
For example,
\textit{"com.apple.security.personal-information.addressbook"} was mapped to \textit{Name,
Physical Address}. In the few cases where there was disagreement, the final
mapping was resolved through mediation and discussion with a third researcher.
The full mapping can be found in Appendix~\ref{appendix:nl_enti}.

Using our mapping, we create an enriched profile for each app
and subsequently compare this against the
label to the actual Nutrition Label to identify discrepancies. We further
analyze the discrepancies based on the individual data items. Each discrepancy
between the generated Label and what the app declared counts as a single
violation.

\noindent \textbf{\circlered{3} Runtime Behavior Analysis.} An app's runtime
behavior provides reliable insights into its data collection practices.
We develop a dynamic analysis approach that evaluates app
behavior across multiple dimensions.

\textbf{\circlered{a} Execution Flow modification.} We follow a two-pronged
approach for observing the app's behavior upon execution: modifying memory at
runtime (required for Frida) and intercepting network traffic via a proxy.

\textbf{\faSearch~\emph{API Hooking.}} Native API hooking is a
technique often used in reverse engineering to redirect the flow of the
execution pointer.  We use Frida, a dynamic instrumentation tool that adopts
a \emph{trampoline} code injection method, wherein it modifies the system's
execution flow so that when a target function is called it redirects the
system to run custom code before returning to the target
function~\cite{frida,Frida:OSDC:2015}. Frida does so by modifying the memory
location of a target function on the fly.  We modify the flow of the target
process by adding logic within two event blocks; (a)~code within an
\texttt{onEnter} block that executes before the target function is executed, and (b)~code
within an \texttt{onLeave} block that executes after the target function has
executed. The modified function returns a callback, including a stacktrace of
the current process, to the aforementioned event blocks, which allows us to
differentiate between first-party and third-party execution. 

Frida's \emph{trampoline} injection method 
adds overhead, especially for frequently called functions. Prior work that
analyzed the cost of hooking system and function calls across platforms found
that Frida can add 128x physical memory overhead and 129x CPU User Time while
instrumenting \mac~APIs~\cite{lopez_survey_2017}). To ensure that our app
instrumentation remains effective despite this overhead, we balance program
execution with logging, by detaching hooks from a target function after
observing 10 calls to it.  While this may result in our system missing certain
details (e.g., an additional instance of a third-party SDK calling a specific
API that has already been logged), it improves our system's efficiency and
reliability.

Reading and modifying processes on macOS requires access to the
\texttt{task\_for\_pid} Mach (IPC) call, which is denied by default by System
Integrity Protection (SIP, an XNU kernel extension). We disable SIP on the test
device by booting it into recovery mode~\cite{integrity_protection}. Disabling
SIP can be detected by apps, which could result in deceptive
apps changing their behavior to avoid detection. As such, our findings
in \S\ref{sec:evaluation} present a lower bound of privacy-invasive behaviors.

\textbf{\emph{Network Proxy.}} Apart from recording relevant API access, we
inspect network traffic to detect the exfiltration of collected data.  We use
the man-in-the-middle proxy (\texttt{mitmproxy})~\cite{mitmproxy}, which
requires installing root certificates on the device so as to capture all
HTTP(S) and DNS requests sent from the evaluated app.  \system parses the
collected network traces to identify instances of collected data being
transferred, along with details of the domains that receive it.

We  also integrate certificate-unpinning via the SSL Kill Switch~\cite{kill_switch} Frida script, which
overrides SSL checking functions in several popular libraries. This script is also
injected before the entry point of the executable allowing us to intercept TLS traffic 
from application launch.
Apps may include customized SSL pinning logic which can be hard to generalize
and bypass thus leading to a subset of network requests not being
intercepted. This is a known challenge, which has affected prior studies as
well~\cite{lalaine}.

\textbf{\circlered{b} User Interaction.} UI automation is performed using a
heuristic-based approach via Apple's Accessibility API. Our script captures the
app state as a nested tree data structure, where each UI element forms a node.
This tree is fetched after each interaction; login pages and forms are detected
via heuristics and populated with synthetic data. The interaction script
follows a weighted semi-random approach to select interactable elements and
execute the selected action.

The tool identifies
elements such as Text fields and generates random text of varying lengths to fill them. 
The tool can also provide apps with inputs from the filesystem via Apple's
Finder App (filepicker), such as picking a media file for a media player
or specifying the output directory.
Our navigation flow covers a wide range of user interactions,
so that our tool can navigate an app more effectively than
basic monkey testing~\cite{exforsys_what_2011}.

\textbf{\faSearch~\emph{Permissions.}} Apple enforces runtime checks on
sensitive resources like the camera, microphone, and filesystem. In such
situations, \mac~prompts the user to provide explicit permissions.  Once granted
(or denied), the system records this choice in a user-specific database that
it queries during future attempts. We ensured that all user permission prompts
that pop-up during interaction are accepted. This has two benefits: (i)~it
ensures that our tool's interaction with the app remains uninterrupted, and
(ii)~it allows us to capture all observed permission prompts by accessing the
app's permission database at the end of its interaction session.

\noindent \textbf{\circlered{4} Privacy Nutrition Label Analysis.}
Subsequently, \system collects the logs generated during the app's execution
and compares them against the app's Nutrition Labels. We consider four perspectives
(marked with~\faSearch~in Figure~\ref{fig:framework}) to an app's data
collection practices. We independently consider each source and develop unique
mappings between its taxonomy and the taxonomy used by Apple's Nutrition
Labels. We first map API calls to the data types which should be declared.
Then for each API call made by the app, we add the mapped labels to create an
\emph{accurate} Nutrition Label, which is the union of all unique data labels.
\system{} then compares the results of the mappings against the actual
Nutrition Labels obtained from the App Store, and flags any discrepancies.

\noindent \textbf{\circlered{5} Extended Tracking Analysis.} While Nutrition
Labels provide an overview of the app's data collection, they do not
accurately capture its tracking capabilities, as they \emph{do not} account
for device fingerprinting (in online tracking this is typically
achieved through browser fingerprinting~\cite{laperdrix2020browser}). This
relies on collecting a wide range of device characteristics like the screen
resolution, model, OS version, GPU vendor, etc. Individual attributes
are not unique but can provide a high-entropy identifier when combined. To
account for such practices, we created a table of attributes using MacOS's
\texttt{system\_profiler} and the test user profile, and mapped it to
individual data items like \texttt{user id} and \texttt{device id.}
We also included coarser device attributes not covered explicitly by \nl.
Our system searches for these attributes in the network traffic.  This is a
well-established strategy for identifying apps exfiltrating data of interest; we
adapted hashing and encoding techniques from prior
work~\cite{englehardt2018,Senol:2022}. We consider multiple hashing 
(e.g.,~MD2/4/5, SHA1/224/256/384/512, MMH3) and encoding (e.g.,~Base16/32/58/64, gzip,
LZ-string) functions. For each identifier of interest, we pre-computed a pool
of potential \emph{multi-layered} hashed and encoded values, which we
search for
in captured traffic including the Request URL,
the body of \texttt{POST} requests, and response bodies.

\section{Experimental Measurements}
\label{sec:evaluation}

In this section, we provide the first measurement study on 
privacy violations in the macOS desktop environment.
We demonstrate our system's capabilities in detecting
application-component discrepancies in bundle-level 
declarations~(\S\ref{subsec:bundle_analysis}), identifying \nl violations via data
exfiltration through application behavior~(\S\ref{subsec:dynamic_analysis}),
and also extend our analysis to device-based fingerprinting capabilities in
\mac applications~(\S\ref{subsec:fp}).

\textbf{Experimental Setup.} We use an M4 Macbook Air to run our experiments,
and create a fresh Apple account to download apps onto the device from the
official Apple App Store. To collect representative data, each app is individually executed for $\sim$5 minutes, a threshold selected 
after exploratory analysis showed no additional violations beyond this point. This is also consistent with prior research on mobile dynamic 
analysis~\cite{arkalakis2024abandon,lalaine} (nonetheless, we note that our results represent a lower bound, which could be further amplified with longer exercise times).
  For apps requiring an account, a
single account was created while all other aspects of the app interaction, such
as logging in and UI clicking, are automated.  After completing the dynamic
analysis, all callback logs are automatically processed to generate  accurate
\nl for each application and to produce an aggregate report including the
violations and discrepancies detected by \system across our dataset.

\textbf{Dataset.} While Google provides detailed ranking and popularity metrics
across app stores (e.g., Google Play Store, Chrome Extension Store), the Apple
App store does not disclose specific statistics or a single global ranking for
macOS or iOS apps. Instead, it publishes category-specific lists of popular
apps that are updated based on proprietary metrics, including recent download
activity.  Since download statistics are not publicly available, we use the
number of user ratings as an indicator of application popularity.

To construct our dataset, we initially assembled a list of all app URLs via the
sitemap\cite{apple_sitemap} of Apple's website, which we used to scrape their
app webpages.  Interestingly, upon closer inspection, we found that Apple's
sitemap only includes iOS and iPad apps while excluding macOS apps.  Moreover,
when a particular macOS app also has an iPad or iOS version, the app's
metrics (such as its rating or number of ratings) correspond to the iPad or iOS
version instead of the macOS version. The macOS version's ratings, which often
differ significantly from the iOS or iPad version, are only available within the
dedicated App Store application.  

To enhance our dataset, we leveraged AppFigures~\cite{appfigures}, a popular
app analytics platform. We scraped the top 250 ranked Mac apps for each of the
20 different categories, merged them and filtered for apps which were designed
for macOS. For each selected app, we additionally scraped their app store
webpage to collect the user rating counts. This resulted in over
1,800+ apps, ranging from highly popular apps such as Microsoft Word (1.7M
ratings) to smaller and less popular applications such as the Visio File Editor
(1 rating). We subsequently excluded ~350 applications that employed anti-analysis techniques,
which prevented dynamic instrumentation via Frida. 
Accordingly, our final dataset (referred to as \texttt{TotalApp}) consists 
of two different subsets:~(i) 500 popular apps based on the
number of user ratings (\texttt{TopApp}), and ~(ii) 500 apps randomly selected from the
remainder of the collected apps (\texttt{RandomApp}). 
\texttt{TopApp} spans a wide range of categories, including Productivity (17\%) Utilities (15\%),
Games (5.7\%), Photo and Video (8.7\%), Business (5.9\%) Music (6.9\%) and Health \& Fitness (2.7\%)
among others. Similarly, \texttt{RandomApp} includes applications from categories such as Productivity (3.3\%),
Utilities (5.7\%), Developer Tools (10\%), Graphics \& Design (8.1\%), Education (5.5\%)
and Business (5.9\%) among other categories.  For each app in the dataset, our
system parses the \texttt{Info.plist} to extract the bundle identifier, and to
also retrieve each application's information and nutrition label values via
iTunes's public API~\cite{itunes}. We also detected that 14.9\% of the
applications restricted functionality by requiring an account, for which we
manually registered a dedicated test account.  In addition, 47 applications in
our dataset offered a free trial associated with our Apple account, which we
activated during dynamic analysis~\cite{charles_what_2024}.

\textbf{Attribution.} As part of our analysis, we are interested in attributing
data access and potential discrepancies to first- or third-party code.
We define first-party origin to  include both application developer provided components and Apple-provided system libraries and frameworks bundled with macOS, as these are required dependencies for all applications. Any other origin is classified as third-party.
For Privacy Manifests, each file is present inside
a parent folder which contains the name of the  associated framework or the SDK.
Since native macOS components do not define manifest files,
we use the location of the manifest file in the app bundle to determine
whether it is first party or third party.

Regarding API hooking  and backtrace attribution, our system captures the
sequence of binaries, dynamic libraries and frameworks involved in each
function call.  These typically include Apple frameworks and system libraries
(e.g., UIKit~\cite{uikit}, CFNetwork~\cite{cfnetwork},
\texttt{libsystem\_pthread.dylib}, and \texttt{libobjc.A.dylib}), the
application’s core binary (e.g., VPN Unlimited~\cite{vpnunlim}), or third-party
libraries and frameworks (e.g., FirebaseCore or
FirebaseCrashlytics~\cite{firebase}). First-party binaries are identified by
matching the binary name to the application name.

The primary challenge in this process, is to identify the native framework and
SDKs, since there is not available documentation for all system components.  To
address this, we curated a reference list using a combination of open-source
GitHub repositories~\cite{macheaders,macsdk}, search engines, and macOS system
file listings.  Any frameworks, SDKs, or binaries not identified as first-party
or native Apple components are classified as third-party.

\subsection{Bundle-Declared Privacy Capabilities}
\label{subsec:bundle_analysis}

\begin{table}[t!]
\centering
\caption{Distribution of apps with discrepancies between their declared
    entitlements and corresponding Nutrition Label data types. (For more
    details see Table~\ref{tab:extended}, Appendix.)}
    \label{tab:entl_discrepancies}
\resizebox{\columnwidth}{!}{%
\begin{tabular}{lccc}
  \toprule
    \textbf{Entitlement} & \textbf{Discr. (\%)} & \textbf{Top. (\%)} & \textbf{Random. (\%)}\\
    \midrule
    \multicolumn{4}{c}{\textbf{com.apple.security}} \\
    \hline
    \rowcolor{lightgray} files.user-selected.read-only & 95\% & 94.1\% & 95.7\%  \\
    personal-information.location & 93.6\% & 93.8\% & 93.4\% \\
    \rowcolor{lightgray} personal-information.addressbook & 93.8\% & 93.9\% & 92.8\% \\
    files.downloads.read-write & 92.3\% & 93.8\% & 89.6\% \\
    \rowcolor{lightgray} personal-information.photos-library & 90.8\% & 90\% & 92.1\% \\
    files.downloads.read-only & 88.8\% & 66.6\% & 100\% \\
    \rowcolor{lightgray} files.user-selected.read-write & 90.7\% & 87.7\% & 94.5\% \\
    device.microphone & 89.4\% & 88.2\% & 90.4\% \\
    \rowcolor{lightgray} assets.pictures.read-write & 87.1\% & 89.4\% & 83.3\% \\
    personal-information.calendars & 86.6\% & 84.7\% & 92.3\% \\
    \rowcolor{lightgray} device.audio-input & 86.1\% & 87.6\% & 84.5\% \\
    device.camera & 84.5\% & 84.5\% & 84.4\% \\
    \rowcolor{lightgray} assets.pictures.read-only & 80.9\% & 75\% & 100\% \\
    temporary-exception.files.home-relative-path.read-only & 68.7\% & 61.5\% & 100\% \\
    \midrule
    \multicolumn{4}{c}{\textbf{com.apple.developer}} \\
    \hline
    healthkit & 100\% & 100\% & 100\% \\
     \rowcolor{lightgray} healthkit.access & 100\% & 100\% & 100\% \\
    ubiquity-kvstore-identifier & 93.6\% & 90.5\% & 100\% \\
     \rowcolor{lightgray} networking.networkextension & 90.3\% & 89.4\% & 91.6\% \\
    ubiquity-container-identifiers & 90.7\% & 87.1\% & 97.3\% \\
     \rowcolor{lightgray} in-app-payments & 66.6\% & 50\% & 100\% \\
    game-center & 65\% & 66.6\% & 64\% \\
     \rowcolor{lightgray} financekit & 50\% & 50\% & N.A. \\
    \midrule
    \multicolumn{4}{c}{\textbf{com.apple}} \\
    \hline
    \rowcolor{lightgray} mail & 100\% & 100\% & 100\% \\
    accounts.appleaccount.fullaccess & 100\% & N.A. & 100\% \\
    \bottomrule
\end{tabular}
}
\end{table}

In the first stage of our analysis, we analyze each application’s bundle to
collect its declared privacy capabilities.  We extract the set of permissions
specified in its \texttt{Entitlements} and parse the associated \texttt{Privacy
Manifest} to determine the declared data-access requirements. 
Both entitlements and manifests
specify these permissions differently (\cref{sec:background}).
While manifests contain developer disclosures about the data collected (similar to \nl),
entitlements are more granular and directly affect the app's runtime privileges.
Declarations are then compared against the corresponding \texttt{Nutrition
Labels} to identify missing data items, i.e., \emph{discrepancies}.

\textbf{Entitlements}. All 1,000 applications analyzed by \system  declare at least one entitlement, consistent with Apple’s development model, which requires applications to be code-signed with a dedicated  entitlement (e.g.,~\texttt{com.apple.security.app-sandbox}) for effective distribution~\cite{apple_app_sandbox}. 
We identified entitlement discrepancies in 82.8\% of the applications (n=\entitlements), where applications declared at least one entitlement that grants access to data not disclosed in their Nutrition Labels.

Table~\ref{tab:entl_discrepancies} presents the proportion of apps where a
declared entitlement’s data type is absent from the \nl. We find
that the majority of apps exhibit high omission rates across all entitlements, with
several reaching 100\%, indicating that no apps containing that entitlement declared the  associated data access in their Nutrition Labels.
In the \texttt{com.apple.security} namespace discrepancies span both file
operations and user data access. The highest omission rates occur for
\texttt{personal-information.addressbook}, \texttt{files.user-selected.read-only}, 
and \texttt{personal-information. location}. The
\texttt{com.apple.developer} namespace exhibits a similar omission pattern,
with complete omissions for sensitive APIs, including health data and contact
information.

Moreover, Apple-related integrations
(e.g.,~\texttt{mail}, and \texttt{appleaccount.fullaccess} 
lack any corresponding disclosure in the \nl,
indicating that omissions persist even for capabilities tied to highly
sensitive, platform-level services. 
Notably, we found that Apple's own
Swift Playground application doesn't declare the \texttt{fullaccess} entitlement,
an Apple-exclusive entitlement with complete access to the user's iCloud, Mail, etc. 
These omissions have a direct and significant impact on user privacy, as users are not aware of the data access capabilities they grant to applications at installation time.

\begin{figure}[t!]
    \centering
    \includegraphics[trim=0 18 0 0cm,clip,width=0.9\columnwidth]{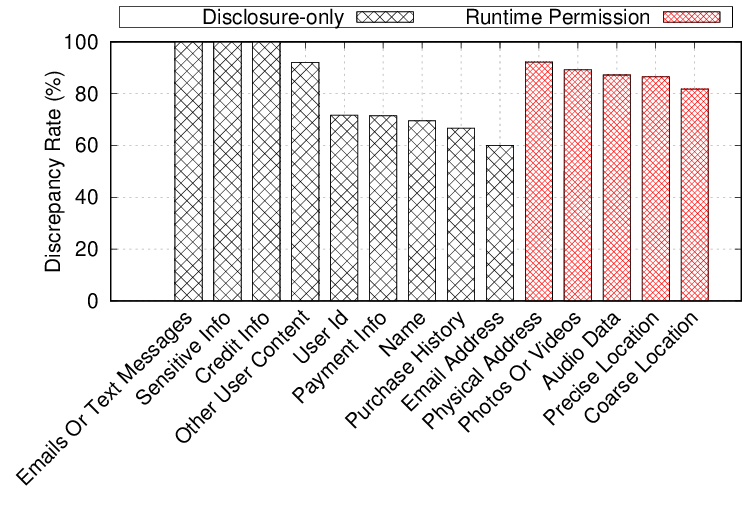}
\caption{Discrepancy rates between declared entitlements and \nls\ data types, grouped by disclosure-only access and TCC-protected runtime permissions.}
    \label{fig:entl_disc_rate}
\end{figure}

We also compare  data labels which are tied to
explicit user consent~\cite{apple_tcc} against data labels which are not. 
As Figure~\ref{fig:entl_disc_rate} shows,  entitlement violations rate remained high in both cases.  
Sensitive Info (Health Data), Emails
and Text Messages, and Credit Info  exhibit 100\% discrepancy,
as none of the apps declared the appropriate entitlements in their \nl.

\textbf{Privacy Manifests.} While entitlements declare system-level
capabilities, privacy manifests disclose data collection at the SDK level.
We extracted \manifesttotal privacy manifests in total from our dataset. Of these,
67 (11.3\%) applications included a first-party or app-wide manifest, which is
generated during the build process, while the majority (88.6\%) originated from
third-party providers. 

Over 70\% of all apps with Privacy Manifests (n=588) did not declare any
\texttt{CollectedDataTypes}. Among the remaining apps (n=168), 135 (80.3\%) apps declared 
\texttt{CollectedDataTypes} in their  manifests that
were absent from the corresponding \nls. 
Overall, discrepancies were detected in 92.3\% of the first-party manifests (n=12) and  in 
78.2\% (n=122) of the third-party manifests.
The discrepancies typically involved fields
such as Other Diagnostic Data, Device ID, and sensitive fields like Coarse
Location and Email Address.

\begin{figure}[t!]
    \centering
    \includegraphics[trim=0 30 0 0cm,clip,width=0.95\columnwidth]{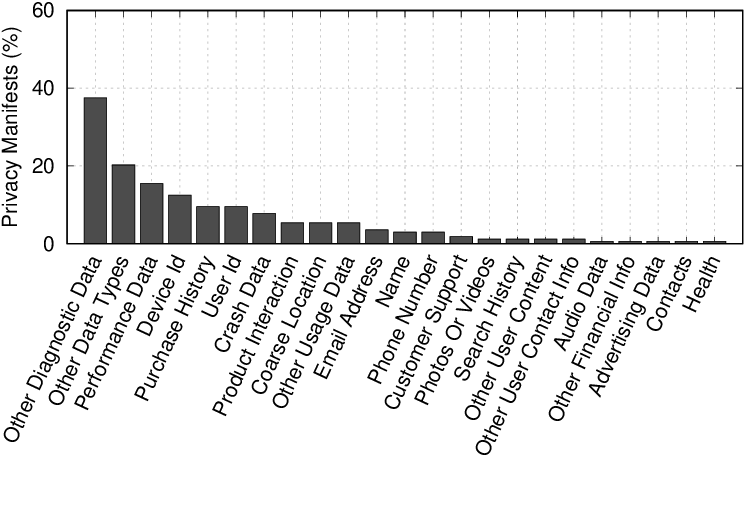}
    \caption{Discrepancies across Nutrition Label data types and privacy
    manifests.}
    \label{fig:manifest_violations}
\end{figure} 

While entitlements allow an app to access specific data types,
manifests disclose the type of data being collected. Each violation
in privacy manifests amounts to a data label being collected without a corresponding disclosure.
Figure~\ref{fig:manifest_violations} illustrates the discrepancies between
privacy manifests and the corresponding Nutrition Labels. The highest affected
type is Other Diagnostic Data, where 37.5\% of manifest declarations lack a
matching Nutrition Label entry. This label represents coarse devices attributes (chipset model, OS versions)
and software telemetry (network or GPU performance). This is a catch-all category
for ``purposes of measuring technical diagnostics related to the app''~\cite{apple_data_tracking_linking}.
Additionally, \nl omissions were detected for Other Data Types (20.2\%),
Device ID (12.5\%), and Purchase History (9.5\%).

Manifest files in SDKs are intended for developers when  integrating third-party components into their applications. The high rate of discrepancies detected by \system{}
indicates that developers often overlook these files.  This may be due to
various factors, such as a lack of awareness of Privacy Manifests or confusion
caused by Nutrition Labels~\cite{LFC_app_study}.  Prior work has also reported
that developers can be confused or misunderstand  the scope of permissions requested by
third-party libraries~\cite{0.1145/3544548.3581060}.  Other cases may exist
due to developers collecting data for debugging and, thus, considering that
data collected for testing purposes does not need to be disclosed.  Finally,
discrepancies could also reflect intentional under-disclosure, as  work has
found that permission abuse can lead to app removal by users~\cite{bonne2017exploring}.

\textbf{Privacy Disclosure Inconsistencies.} Overall, the divergence between
\nls and bundle declarations reveals systematic omissions.
This misalignment enables applications to collect user
and device data without accurate disclosure, thereby preventing users from
understanding the apps' privacy practices. The consistently high
omission rates across major functionalities reveal a structural weakness in
macOS’s verification process, which detects the presence of a Nutrition Label
but does not verify its consistency with declared capabilities.  Given the
scale and compliance obligations of major third-party services, these omissions
demonstrate that the misalignment extends beyond application developers and
further weakens the reliability of Apple's privacy label mechanism.

\subsection{Behavior-based Violations}
\label{subsec:dynamic_analysis}

Next, we analyze application runtime behavior to detect \emph{privacy
violations}, defined as cases in which an app accesses or transmits user or
device data without a corresponding \nl declaration. 
We compare their data flows to declared capabilities, to detect
non-disclosed data collection and transmission, and reveal behavior-based
inconsistencies beyond the bundle-level analysis.

\begin{figure}[t!]
    \centering
  \includegraphics[width=0.8\columnwidth]{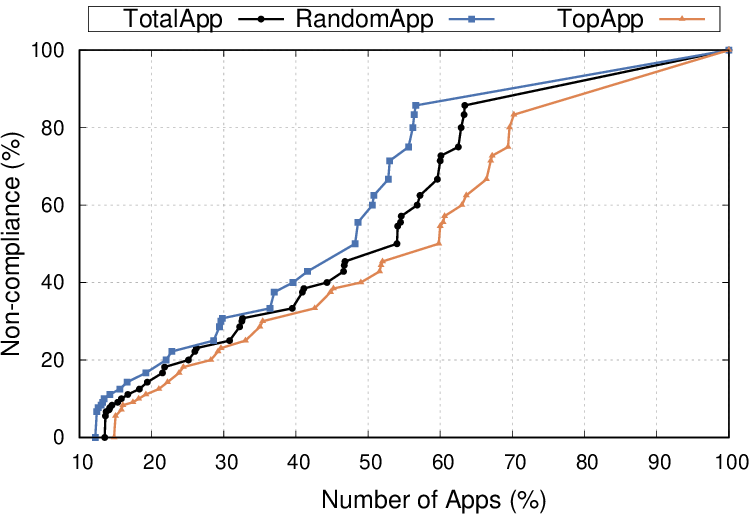}
    \caption{Degree of \nls non-compliance across evaluated applications. }
  \label{fig:compliance}
  \end{figure}

\textbf{Runtime Traces.}
We applied our dynamic analysis pipeline (\S\ref{sec:implementation})
to \texttt{TotalApp}, collecting execution traces, native and
third-party API calls, and associated data.  \system{} then maps the runtime
behaviors to their corresponding \nl categories, allowing the direct comparison
to their reported disclosures.  We evaluated app developers' adherence to App Store
requirements for declaring data collection practices to end users, by comparing
the Nutrition Labels with the data collection observed through the hooking of Native API Calls.
To quantify the extent of \emph{non-compliance} we calculate
the following metric:

$\text{Non-compliance} = 1 - \frac{A \cap B}{B}$ \\
\indent A:= Set of Data Types declared within Nutrition Labels.\\
\indent B:= Set of Data Types observed from Native API Calls.\\
\noindent Figure~\ref{fig:compliance} illustrates the distribution of
\emph{non-compliance} for the apps in our dataset. 
Each undeclared data type (e.g., Name or User ID) is counted as a single violation, regardless of whether it is accessed by one or multiple API calls.

Approximately 15\% of the apps were fully consistent with their \nls,
achieving complete compliance and accurate privacy disclosure. Prominent apps
among these include Microsoft Excel, One note and Outlook. While these apps
collect sensitive identifiers like Device ID, Email address, Coarse Location
and Contact Data, they accurately disclose it in their \nl. Nevertheless, the fact that  only
15\% of the apps achieving full compliance is a concerning finding and
comparable to the extent of non-compliance reported for mobile
apps~\cite{arkalakis2024abandon}.

 \begin{figure}[t!]
  \centering
  \includegraphics[width=0.9\columnwidth]{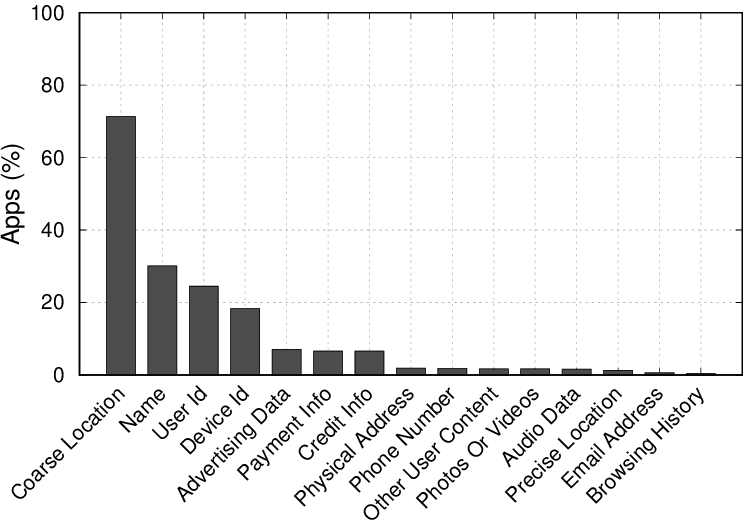}
    \caption{Distribution of violations by Nutrition Data Labels.}
  \label{fig:violation_type}
\end{figure}

Moreover,  60\% of the apps omitted half the data types that
they collect from their Nutrition Labels, and approximately 38\% did not
declare \emph{any} of the data types, resulting in total non-compliance. While
most of these cases are less popular apps, the popular LetsVPN
app~\cite{letsvpn}\footnote{While the Apple App Store does not provide download
numbers, indicatively, LetsVPN has 5M+ downloads in the Android Play Store.} 
declares in its \nl that ``No Data is Collected``. To make matters worse, LetsVPN's privacy policy~\cite{letsvpnprivacy} which is publicly available in the app's App Store page, states that the app collects Device ID, Purchase Information among other data types.
Under Apple’s Nutrition Label guidelines, this type of data collection is required to be disclosed. 
\system further detected that LetsVPN accesses User ID, Device ID, and Coarse Location at runtime, indicating a discrepancy between declared and observed data practices.
In contrast, we find that widely used apps (e.g., WhatsApp) are more likely to provide accurate Nutrition Labels.

Notably, we detect substantial variance in the non-compliance rate between
\texttt{TopApp} and \texttt{RandomApp}.
A larger fraction of apps in the randomly
selected dataset are non-compliant compared to popular apps.
Nearly 45\% of apps in the random dataset are fully
non-compliant compared to 30\% for popular apps. 
One potential factor contributing to the higher non-compliance among less
popular applications is differences in available resources. Popular
applications are typically developed by larger organizations with dedicated
teams, which allows higher attention to privacy compliance compared to less
popular applications.  Our results demonstrate that runtime behaviors
diverge significantly from declared privacy practices, indicating that Apple’s
disclosure mechanism is neither effectively enforced nor consistently adopted
by developers.

\emph{Data type Violations.} To quantify the impact of runtime behaviors on the
accuracy of \nls, we mapped omitted data types for each app and
measured their prevalence. As shown in Figure~\ref{fig:violation_type}, the
most frequently undisclosed item was Coarse Location, detected in approximately
70\% of the apps and primarily exposed through Apple’s \texttt{NSUserDefaults}
API, which exposes the nearest major city. 
This value is set by macOS when selecting the system's timezone.
Other categories
of PII, including Name (30.1\%), User ID (24.5\%), and
Device ID (18\%), were also frequently collected without disclosure.

Since apps access combinations of identifiers and data
in parallel, violations do not occur independently.
The most common
combination involved was User ID, Name, and Coarse Location, while another frequent
pattern involved simultaneous collection of Device ID and Coarse Location.
These overlapping violations increase privacy risks as they can be leveraged to cross-link
user identities across devices and online contexts.

\begin{figure}[t!]
  \centering
  \includegraphics[width=0.8\columnwidth]{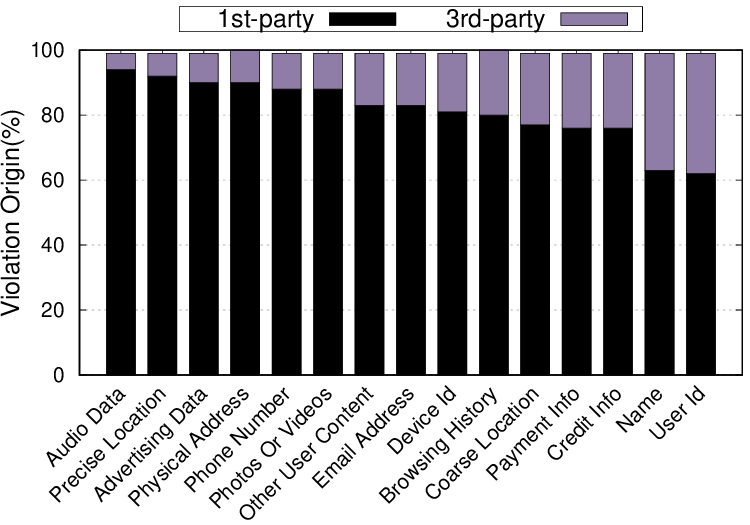}
    \caption{Breakdown of violations by attribution to first or third party.}
 \label{fig:violation_origin}
     \end{figure}

\emph{Violation Origin.} We further attribute  API accesses to first- or third-party based
on the Frida backtrace. This backtrace captures the stack at the time of
execution, and maps the memory locations to the relevant code that
resulted in the call. To associate undisclosed data access to the originating
party, we apply the attribution process described previously. Each mismatch between 
an API call and data label item is considered a separate violation. 

As shown in Figure~\ref{fig:violation_origin}, the majority of violations originates from
first-party behaviors.
This finding contrasts with prior findings for Android applications, where the majority of the violations were attributed to third-party code~\cite{arkalakis2024abandon}.
In more detail, for multiple data types (e.g.,~Audio Data, Precise Location,
Browsing History) the majority of violations was attributed to the
first-party. In contrast, identifiers such as User ID, Device ID, and Name are
more evenly collected by both parties, with up to 40\% of violations linked to external SDKs and libraries.

We detect a high number of first-party \nl violations, in contrast to privacy
manifests where third-party violations dominate.  This discrepancy is driven from the broader availability of third-party manifests, which results in a higher
share of violations originating from external SDKs.  Notably,  one-third
of apps exhibits third-party \nl violations, a proportion comparable to those with third-party manifest violations.  Overall, both first- and third-party
origins contribute to privacy violations, although this distinction is less
pronounced in the manifest analysis.

\begin{figure}[t!]
    \centering
  \includegraphics[width=0.8\columnwidth]{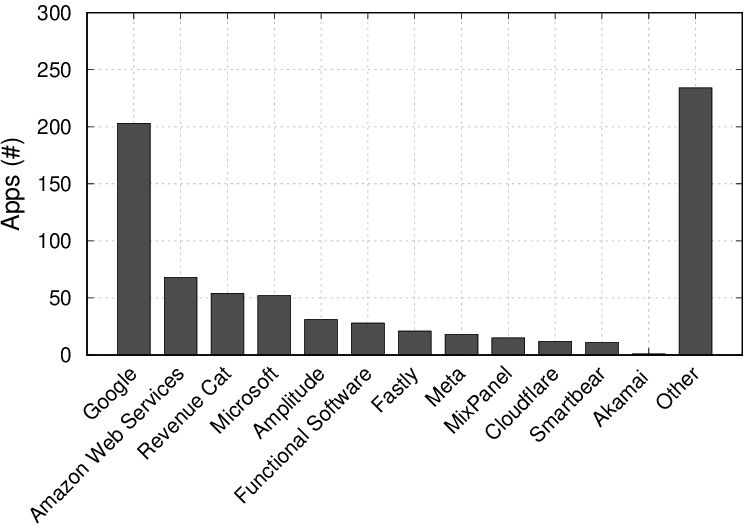}
    \caption{Top third-party entities present in apps. }
  \label{fig:by_company}
\end{figure}
\textbf{Third-Party Entities.} While detecting undeclared access at runtime
establishes that apps collect sensitive data without disclosure, 
it does not determine whether data is  directly exfiltrated beyond the
application boundaries. To assess whether API access also leads to data
leakage, we analyze the network traffic generated during the app's execution 
and categorize the prevalence of third-party domains being contacted.
Figure~\ref{fig:by_company} illustrates the distribution of entities contacted by all apps across the dataset. 
Each domain was categorized using the
TrackerDB~\cite{trackerdb} and Disconnect tracker lists~\cite{disconnect},
and then mapped to their parent service provider based on the \texttt{eTLD+1}.  
The most common services originate from  Google, reflecting their
major roles as SDK  and service providers. Amazon Web Services is also
prevalent, operating primarily as a hosting infrastructure. The ``Other''
category captures a long tail of third-party domains, each associated
with a smaller number of apps. We note that these patterns mirror those
measured on web and mobile platforms, where tracking/analytics/infrastructure
is concentrated to a limited number of major providers~\cite{dambra2022sally}.

\emph{Data Exfiltration.} We next assess whether the captures network communications, transmits sensitive user or device data.
In total, 497 applications transmitted user and  device information over the network. 
Among these, 204 applications transmitted identifiers explicitly protected by the \nls, and 62 of those did not disclose the corresponding data types.
The remaining apps transmitted coarser, fingerprint-related attributes, such as kernel or OS version, Mac model identifier, and screen resolution.
Table~\ref{tab:exfil_app} reports the breakdown of undisclosed data types collected, and the top domains receiving the data.
Across all data types, the most commonly contacted entities were Google (\texttt{googleapis.com}, \texttt{1e100.net}), Amazon (\texttt{amazonaws.com}, \texttt{cloudfront.net}), and Facebook (\texttt{fbcdn.net}, \texttt{facebook.com}). 
These entities primarily provide advertising, analytics, or infrastructure services, highlighting the privacy implications of undisclosed data transmission by desktop apps.

\subsection{Extended Fingerprinting Analysis}
\label{subsec:fp}

In the final part of our study, we present a taxonomy of all device
identifiers we encountered during our analysis. This  analysis is independent of the current design of Nutrition Labels and
instead, we consider all device identifier and related attributes transmitted
off-device.  We analyze their prevalence in app network traffic and categorize
them by the type of information they reveal, demonstrating how third parties
can leverage them for augmenting device fingerprinting.  The identifier list
was generated via the \texttt{system\_profiler} command on macOS, which provides
a highly detailed technical overview of a device's hardware capabilities.

Table~\ref{tab:device_ext} lists the identifiers detected, representative
values, and their prevalence across apps.  Several device-specific
identifiers (e.g.,~\emph{System Version} and \emph{Kernel Version}) are
disclosed by default through Apple’s networking stack, as they are embedded in
the user-agent string automatically generated by \texttt{CFNetwork}, the default
networking library~\cite{cfnetwork}. Interestingly, the version value increments with each operating system update (e.g., macOS
15.6, Darwin 24.5.0), allowing trackers to classify the devices by patch level.
This leakage extends the fingerprinting surface beyond Apple's declared \texttt{Device ID} category, introducing a persistent, fine-grained feature for tracking.
Additional identifiers expose detailed device characteristics and
configuration information. The \texttt{Model Identifier} directly encodes hardware
specifications; for example, \texttt{Mac16,12} denotes the 13-inch MacBook Air
with an M4 chipset~\cite{modelid}.  We detected  this identifier in
the network traffic of 222 apps (44\%), representing a significant fraction
of our dataset.  
Apps also transmitted screen resolutions attributes that
are considered as high-entropy for fingerprinting within 
browsers~\cite{device_finger}. While native resolution can also be
inferred from the model identifier, the HiDPI value varies with user
configuration, introducing a distinct fingerprinting modality that expands the
range of device-specific attributes available to third parties.
Furthermore, identifiers exposed through Apple's Graphics API, Metal~\cite{apple_metal}, also
introduce an independent hardware feature, further enriching the fingerprinting
surface. Finally, we detected a specific combination of identifiers (``Darwin 12.3.0" , ``15.3.2", ``M4", ``<REDACTED\_CITY>")  in 72 apps' network traffic, directly revealing the device and execution environment.

\textbf{Summary.} Our analysis highlights the various shortcomings in the current design of
Apple Nutrition Labels. In its current state, the Nutrition Label framework is insufficient for 
both developers and end users. Developers must navigate the complex
compliance system encompassing entitlements, privacy manifests, and Nutrition Labels.
In addition, the lack of \nl verification during App submission also means that users encounter incomplete 
or incorrect labels, undermining their trust in the system.
Furthermore, given the proliferation of device fingerprinting
for user tracking and risk-based authentication, and the extent of 
device-identifier exfiltration detected by \system{}, the absence of a dedicated fingerprinting
category in Nutrition Labels is a significant oversight. While Apple has acknowledged 
the fingerprinting potential of certain system APIs\cite{apple_required_reasons}
and requires developers to disclose why they are used, this disclosure is 
relegated to the privacy manifests which most end users will neither read nor encounter.
A ground-up redesign of Nutrition Labels
that focuses on simplifying Apple's data collection nomenclature, and extending
it to include device fingerprinting, will provide a more comprehensive and
impactful privacy disclosure mechanism.

\begin{table}[t!]
\centering
\caption{Applications exfiltrating sensitive data in violation of \nl, and the corresponding third-party domains.}
\label{tab:exfil_app}
\resizebox{1\columnwidth}{!}{%
\begin{tabular}{l|c|c}
    \toprule
    \textbf{Data Type} & \textbf{Apps} & \textbf{Domains}\\
    \hline
     \rowcolor{lightgray}
      & & cloudfront.net, akamaitechnologies.com,\\
     \rowcolor{lightgray}
     \multirow{-2}{*}{Name} & \multirow{-2}{*}{44} & fbcdn.net, amazonaws.com, 1e100.net\\
    Coarse & \multirow{2}{*}{124} & googleapis.com, 1e100.net,\\ 
     Location & & amazonaws.com, facebook.com, braze.com \\
     \rowcolor{lightgray}User ID          & 10  & cloudfront.net, amazonaws.com, 1e100.net \\
     Device ID        & 16   &  google.com, unity3d.com, icandesignapp.com \\
     \rowcolor{lightgray}Email & 7 & 1e100.net, amazonaws.com \\
    \bottomrule
\end{tabular}
    }
\end{table}

\begin{table}[t!]
\centering
\caption{Access of fingerprintable device attributes.}
\label{tab:device_ext}
\footnotesize
\begin{tabular}{l|c|c}
   \toprule
    \textbf{Device Attribute} & \textbf{Description} & \textbf{Apps} \\
    \hline
     System Version & macOS 15.3.2 & 362 \\
    \rowcolor{lightgray} Kernel Version & Darwin 24.3.0 & 408 \\
     Chipset Model & Apple M4 & 241  \\
     \rowcolor{lightgray}Location &  \texttt{Redacted}& 145 \\
     Model Identifier &  Mac16,12  & 222 \\
     \rowcolor{lightgray}Resolution(HiDPI)  & 1470 x 956 Retina & 145 \\
      Email  & \texttt{Redacted} & 24 \\
    \rowcolor{lightgray} Display Type & Liquid Retina XDR Display & 147 \\
     Name &  \texttt{Redacted} & 59 \\
     \rowcolor{lightgray}User Name & \texttt{Redacted} & 59 \\
    Boot Volume & Macintosh HD & 78 \\
     \rowcolor{lightgray}Hardware UUID & \texttt{Redacted} & 21 \\
     Metal Support &  Metal 3 & 54 \\
     \rowcolor{lightgray}Computer Name &  \texttt{Redacted}'s MacBook Pro & 53 \\
     Bus & Built-In & 18 \\
     \rowcolor{lightgray}Serial Number (system) & \texttt{Redacted} & 11 \\
    Hardware & AirPort & 21 \\
      \rowcolor{lightgray}Domain Name & 10.0.5.1 & 3 \\
     MAC Address & \texttt{Redacted} & 6 \\
     \rowcolor{lightgray}IPv4 Addresses & 10.0.5.14 & 2 \\

   \bottomrule

\end{tabular}
\end{table}

\section{Mitigating Disclosure Discrepancies}
\label{sec:mitigations}

A key takeaway from our measurement study and data analysis is that disclosures
cannot rely on developer self-reporting, and effective mitigations require
replacing this implicit-trust-based model with centralized verification. \system
demonstrates that automated analysis can generate \emph{reliable} and
\emph{accurate} disclosures by cross-referencing bundle components. These
audits can be integrated directly into build and deployment pipelines. For
instance, \nls can be pre-populated from SDK manifests embedded in the app,
reducing redundancy and minimizing omissions.~\texttt{Xcode} already has the
ability to combine all third-party manifest files into an app-wide manifest
file. Yet, developers are currently required to manually fill out \nls.

\textbf{Platform changes.} At the platform level, enforcement must
combine static and dynamic validation. The submission pipeline should
incorporate automated cross-checks between \nls, entitlements, and manifests,
rejecting apps with inconsistent disclosures.  \nls can be partially
auto-generated from bundle artifacts (e.g.,~SDK manifests) at build time,
ensuring that developer-supplied disclosures cannot fall below a verified
baseline. \texttt{Xcode} should partially autofill the Nutrition Label component,
and interface with App Store Connect to ensure the presence of baseline
disclosures during submission. Finally, runtime enforcement should be extended
to detect undeclared access to protected APIs. These measures must be
integrated by Apple across deployment and runtime mechanisms to replace the
current fragmented model.


\noindent\textbf{Unified Disclosure and Verification.} Apple uses a
combination of property list declarations and user-enforced runtime permissions
to regulate access to sensitive data, while other privacy-related declarations
(e.g., nutrition labels) remain unverified. This fragmented approach creates
opportunities for inconsistencies and omissions. 

We provide an example that illustrates the problematic nature of the current
design: imagine a developer that wants to create an app that can record
videos. The app would need to use multiple
\emph{Native APIs} in their code; \texttt{AVCaptureSession}
begins an event during which the user takes a picture. Since Apple considers
access to this functionality sensitive, the developer would need to request the
user for a \emph{runtime permission} to access the camera via a different
object, \texttt{AVCaptureDevice}, whose property \texttt{requestAccess}, is
used to trigger a system prompt that is displayed to the user. However, when
\mac~observes this API call, it looks for an entry in the \emph{Entitlements}
file, \texttt{com.apple.security.device.camera}, which is a static declaration
from the developer that the app will access the camera. In the absence of this
entitlement, the developer's code )
will not prompt the user and hence will not receive device permissions.
Additionally, if the developer wants to list their app on the App Store, they
need to create a \emph{privacy manifest} file, and include a key,
\texttt{NSPrivacyCollectedDataTypePhotosorVideos}, along with details about how
the data is used and handled by the app. While the manifest entry does not
affect runtime enforcement, it is used by App Store reviewers to approve the
app for listing. Finally, the developer must also declare the
\texttt{Photos \& Videos} type in the \nl, under the category \texttt{User Content},
to inform users that the app accesses their photos or videos.
Listing~\ref{lst:example-camera-access} (Appendix~\ref{appendix:unified}) provides an example
snippet of Swift code a developer might use for this workflow.

The developer in this example creates redundant declarations, each of which uses
different terminology. Alternatively, Apple could consolidate these components into a single
declaration that is used for both enforcement and user disclosure. The
developer could declare in a single file that the app accesses the
camera to record videos, and this declaration would be used by the OS to
enforce access restrictions, by App Store reviewers to approve the app for
listing, and by users to understand the app's data collection practices.
Considering \texttt{Xcode}'s existing capabilities to populate and consolidate
\emph{entitlements} and \emph{privacy manifest} files, we propose that a
similar approach should consolidate all data access into a single declaration.
We provide an example declaration in Appendix~\ref{appendix:unified} for this
specific example, illustrating our proposed unified approach. This strategy 
would streamline the disclosure process for developers by eliminating
redundancy, minimizing accidental omissions, and enhancing the reliability of
disclosures, while also facilitating a platform-level automated verification 
process.

\section{Discussion and Limitations}
\label{sec:discussion}

While \system~extensively evaluates the runtime behavior of \mac~apps, its
analysis has the following limitations.  \system~uses a custom UI-exercising
tool to interact with apps, but these interactions are not comprehensive, i.e.,
we do not evaluate \emph{all} possible app states or user inputs. As such, our
system may miss certain privacy-invasive behaviors and disclosure violations.
Moreover, our analysis depends on taxonomical and functional overlaps with
Nutrition Labels which can only partially be addressed via Entitlements and
Runtime Permissions. 
Due to the nature of our classification and the absence of platform documentation, a subset of APIs could not be reliably classified into specific label categories.
Our results, therefore, offer a \emph{lower
bound} of data collection, which may be even more extensive than what our study
demonstrates.  Finally, \system~reports on observed data collection but cannot
determine the purpose or use of the collected data, nor can it identify backend
handling practices (e.g., anonymization). These are important for a complete
assessment of app privacy, but can only be performed in collaboration with app
and SDK developers.

\section{Related Work}
\label{sec:related}

Here we discuss pertinent prior work, focusing mainly on research addressing
privacy labels, and desktop platforms.

\noindent\textbf{Privacy Nutrition Labels.} Kelley et al.~\cite{kelley-nutrition-2009}
first proposed the idea of creating a nutrition label for apps similar
to the mandated Food Nutrition labels governed by the FDA. A privacy
nutrition label was designed as a simple and universal disclosure that would
accurately indicate the app's data collection practices. This
would allow a user to understand the data they were sharing and dictate their
decisions about using an app. The idea gained traction, which
led Apple and subsequently Google to implement a version of this nutrition
label calling them App Nutrition Labels~\cite{apple_nutrition_labels} and Data
Safety Labels~\cite{datasafety}, respectively. While initially introduced for
iOS, Apple recently rolled out Nutrition Labels for macOS and made it mandatory
for developers to include one for their desktop apps.  Subsequently,
several dimensions of Nutrition Labels have been studied, including usability
from the end-user perspective~\cite{zhang2022usable}, developer
challenges~\cite{lfc}, and large-scale analyses~\cite{LFC_app_study}.

Closer to our work are studies focusing on compliance. Ali et
al.~\cite{ali2024honesty} compared labels with privacy policies and found
significant divergences across iOS apps. Kollnig et
al.~\cite{kollnig-goodbye-2022} demonstrated that iOS apps used
fingerprinting-derived identifiers for tracking, found evidence of Apple
engaging in tracking, and detected that discrepancies were widespread in less
popular apps, a finding also verified by our research.  Arkalakis et al.~\cite{arkalakis2024abandon} conducted a dynamic analysis of Android apps to
analyze their data collection practices. That study also correlated apps'
dynamic behavior, including API accesses and network traffic, with the apps'
Data Safety Section. Their research uncovered a high degree of non-compliance
with the apps' stated data disclosure policies. While that study follows a
similar methodology (albeit in a vastly different software ecosystem), it focuses on
apps' dynamic behavior depending on users' runtime consent. On the other hand,
our research focuses on the unique aspects of the macOS ecosystem, including
identifying system resources that are not sufficiently tracked through
entitlements, TCC and Gatekeeper, yet can be leveraged by trackers for uniquely
identifying users. Another relevant study~\cite{lalaine} examined iOS apps’
compliance with their Nutrition Labels using dynamic analysis and network
interception. Their dynamic analysis system, Lalaine, detected 67\% of apps to be non-compliant
compared to our 85\%. Their findings are consistent with ours in identifying User ID, Device ID and
Location data as the most frequent non-compliant categories. \system 
analyzes additional app and system level components, including privacy manifests, entitlements, and fingerprintable
identifiers, thus offering a more comprehensive assessment of privacy violations.

\noindent\textbf{Desktop Apps.} Recent research on desktop apps has
focused on the security properties of cross-platform apps.  Carettoni developed
a static analysis tool, Electronegativity, that identifies potential security
vulnerabilities in Electron apps~\cite{Carettoni:2017}.  Xiao et
al.~\cite{xiao2022understanding} found concerning security practices adopted by
640 apps when evaluated with Electronegativity.  They then developed a dynamic
analysis tool to evaluate Electron apps and identify sensitive data flows that
can be exploited to perform Remote Code Execution (RCE) attacks. Similarly, Ali
et al.~\cite{ali2024rise} proposed a dynamic analysis tool, Inspectron, that
runs apps against an instrumented Electron framework and monitors their runtime
behavior to identify potential security vulnerabilities. Jin et
al.~\cite{jin2023security} developed a framework that uses a DOM-tree-based
analysis to evaluate and identify unintended modifications in Electron apps.
Yang and Yue~\cite{yang2020comparative} compared tracking practices between
mobile and desktop environments; however, for the desktop component of their
research, they studied websites that offer a desktop version of their webpage,
they did not study dedicated desktop applications. Blochberger et
al.~\cite{blochberger2019state} compared the deployment and configuration of
sandboxing within macOS applications installed through Apple's official store
to that of applications retrieved from a third-party app store and found that
89\% of apps distributed through the third-party app store did not use
sandboxing, thereby exposing users to risk.

\noindent\textbf{Summary.} To the best of our knowledge, no prior work has
systematically investigated how tracking and fingerprinting techniques are
deployed within desktop apps. This lack of attention is notable given
the research on tracking in web and mobile ecosystems, 
leaving a gap in our understanding of privacy risks in desktop environments. By
presenting the first empirical exploration of privacy-invasive practices in a
desktop platform, our study directly addresses this gap and helps complete the
broader picture of online tracking across different platforms. In addition, we conduct
the first systematic evaluation of macOS apps’ compliance with their
reported Privacy Nutrition Labels. This dual contribution allows us to assess
not only the presence of tracking and fingerprinting behaviors, but also the
extent to which apps adhere to their self-disclosed data practices.
Taken together, our findings provide a more comprehensive picture of the
shortcomings and challenges associated with both data access control mechanisms
and data practice disclosure frameworks in one of the most widely used desktop
operating systems.

\section{Conclusion}


Users have become considerably more aware, and concerned, about online
tracking and other forms of privacy loss, and a rich body of academic
research has explored multiple dimensions of privacy loss within the web,
mobile, and smart device ecosystems. However, despite the continuing widespread
use of desktops, no prior research has explored how desktop apps can
impact users' privacy. To address that gap, we conducted the
first, to our knowledge, systematic evaluation of data practices and violations
by desktop apps. We focused on macOS, due to its general perception as
a more secure and private OS compared to Windows, as well as its recent
adoption of Privacy Nutrition Labels. To that end, we developed \system, an
analysis framework and pipeline that consolidates diverse data sources by
statically analyzing app components and dynamically instrumenting and
exercising app.  Using our system we were able to detect privacy
violations and disclosure discrepancies across a wide selection of popular
apps. Our in-depth analysis of these violations, reveals that known tracking
entities often surreptitiously exfiltrate unique user and device identifiers
that can deanonymize concurrent private browsing sessions, and other sensitive
data that can be used to significantly enrich user's profiles. Our research
calls for more streamlined nutrition-label creation mechanisms, and improved
vetting processes within Apple's App store, to mitigate the significant privacy
risk posed by macOS apps.

\section*{Acknowledgments}\label{sec:acks}

We would like to thank the anonymous reviewers and shepherd for their valuable feedback.
This  work  was supported  by the National Science Foundation under grants
CNS-2211574 and CNS-2143363.  Any opinions, findings, conclusions, or
recommendations expressed herein are those of the authors, and do not
necessarily reflect those of the NSF.

\section*{Ethical Considerations}
\label{sec:ethics}

We conducted this research with strict adherence to ethical guidelines, and
upheld principles in accordance with the Menlo report.

\noindent \textbf{Purpose and Impact of Research.} The primary goal of this
work is to understand the data collection practices of desktop applications and
their resulting implications for the privacy of \emph{end users}.
We show that desktop environments provide rich data collection capabilities, but offer
few guidelines to assist app developers. Our findings can
inform the design of more effective privacy controls and tools for desktop
environments, as well as guide future research in this space.

\noindent \textbf{Beneficence.} At every step of this research project, we
proactively evaluated our findings and its impact on all stakeholders involved,
including \emph{Apple} (especially, reviewers of apps on the App Store),
\emph{app developers}, \emph{developers of third-party SDKs}, and importantly,
\emph{end users}. We highlight their interactions with various studied
components in Table~\ref{tab:threat_model_table} in the Appendix. Below, 
we outline the specific considerations and impact on each stakeholder.

\emph{Apple.} We gathered our dataset of apps from the official Apple App Store.
We fetched individual app bundles and associated metadata one at a time,
ensuring that our requests did not overwhelm the App Store infrastructure. 
Our work demonstrates shortcomings in
the App Store's review process and highlights a need for more comprehensive
verification of submissions.

\emph{App Developers.} All apps and associated metadata that we
analyzed in this study are publicly available on the Apple App Store. 
Our research highlights that developers
are held responsible for a large array of app components in their app, each of
which have partially overlapping semantics and share functional similarities
with each other. Our findings show that these components can result in
inconsistent privacy practices, and can lead to developers including
conflicting privacy declarations for end users to view on the App Store.

\emph{Third-party SDK Developers.} Our analysis highlights the privacy
practices of apps, with all included SDKs, as a whole. We recognize that
developers of third-party SDKs have limited control over how their SDKs are
incorporated within apps, and that the SDKs' data collection practices can vary
based on features accessed by the app itself (i.e., corresponding Native API
calls). We clarify in \S~\ref{sec:background} that the responsibility for
accuracy of submissions lays with the app developers, and point out that SDK
developers can ease the process by providing their own privacy manifest files.
Our proposed system can help SDK developers understand the real-world use of
their SDKs, and better support their clients (i.e., app developers) through
enhanced documentation and support with communicating privacy practices.

\emph{End Users.} The privacy implications of our work primarily
affect end users that download and use apps from the App Store. Our findings
show that end users that rely on the Privacy Nutrition Labels on the App Store
can be misinformed of an app's actual behavior once downloaded. Our work aims
to raise awareness about these issues and provide insights into how apps
collect and use data, thereby empowering end users to make informed decisions
about their privacy.

\noindent\textbf{Respect for persons.} We note that our research does not
directly involve human subjects and informed consent procedures are not
applicable to our research. Our work does not cause any additional risk to end
users, developers, or Apple itself.
Our work informs users, researchers, and regulators thereby
improving transparency and accountability. The study relies exclusively on
empirical measurement of application behavior and analysis of publicly
available disclosures, without interacting with end users or collecting personal
data. While our results may reflect negatively on certain applications or
platform practices, we believe that responsible disclosure of measurement-based
evidence serves the public interest and aligns with established norms in
security and privacy research. Overall, the expected benefits of increasing
awareness and enabling better-informed decisions about privacy risks outweigh
any minimal potential downsides, justifying the ethical conduct of this
research.

\noindent \textbf{Responsible Disclosure.} Upon completion of our analysis, we
provided Apple a detailed
description of our findings, including a list of the apps with discrepancies in
their data practice disclosures. We have received notification that our reported
issues were reproduced and are currently being investigated.


\section*{Open Science}
\label{sec:openscience}

To promote transparency and to facilitate future research, we have made research artifacts available on Github ~\cite{coderepo}.

\bibliographystyle{ACM-Reference-Format}
\bibliography{paper}

\appendix

\section{Privacy Nutrition Example}
\label{appendix:nutrition}

\begin{figure}[t!]
\centering
\includegraphics[width=1.0\columnwidth]{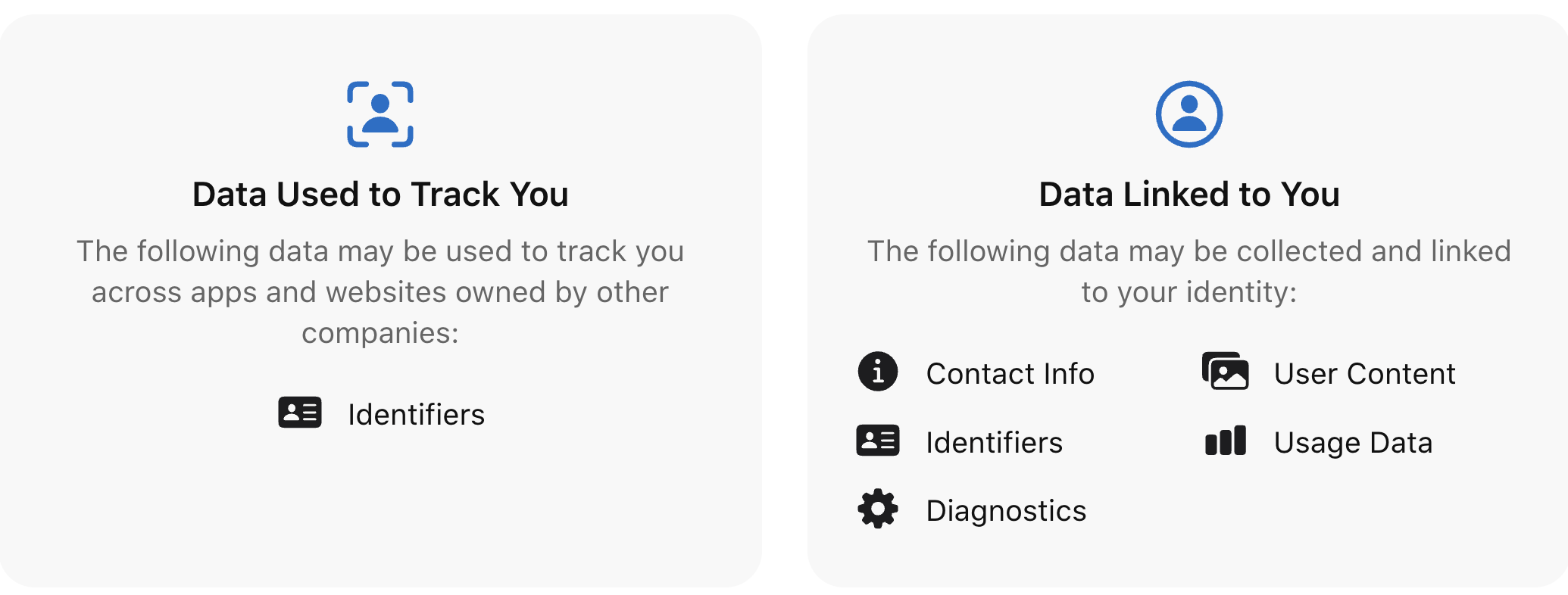}
\caption{The Privacy Nutrition Label of CapCut.}
\label{fig:capcut}
\end{figure}

Figure~\ref{fig:capcut} illustrates the Nutrition Label of a popular video app,
CapCut~\cite{capcut}. The app declares the collection of device identifiers,
user content, and interaction data, and classifies them as ``Linked to You'' or
``Used to Track you''. While this separation increases disclosure granularity,
users face challenges interpreting such information without familiarity with
Apple's labeling system, highlighting a broader limitation of this labelling
approach~\cite{zhang2022usable}. However, while a simpler label listing only
the collected data items may be easier to interpret, it might not accurately
capture applications' data collection practices~\cite{LFC_app_study}.

\section{\mac Privacy Mechanisms}
\label{appendix:capabilities}
\begin{table*}[h!]

\caption{Overview of privacy-relevant components studied in this work with
insights into their relationships and attributes. \textbf{Overlap}
(\faList*~/~\faList*[regular]~= complete/partial Taxonomy Overlap,
\faSun~/~\faSun[regular]~= complete/partial Functionality Overlap) shows
the relationship between components. \textbf{Data Collection \& Handling}
(\faLink~= Data Linking or Anonymization, \faLightbulb~= Purpose,
\faLayerGroup~= Data Types or Categories) indicates the app's data
collection practices that can be gathered from analyzing the component.
\textbf{User Tracking \& Fingerprinting} (\faExclamationTriangle~= Apple's
pre-defined Sensitive APIs, \faFingerprint~= Extended Fingerprinting)
indicates the app's privacy-invasive practices that can be gathered from
analyzing the component. \textbf{Access \& Interaction} (\faPenNib~=
Creation, \faCheckCircle~= Verification, \faEye~= Observation) shows how
each entity accesses the corresponding component.} 
\centering
\label{tab:threat_model_table}
\resizebox{\textwidth}{!}{
\begin{tabular}{@{}l|c|c|c|c|c|c|c|cccc@{}}
\toprule
\multirow{3}{*}{\textbf{Component}} & \multicolumn{5}{c|}{\textbf{Overlap}} & \multicolumn{1}{c|}{\textbf{Data}} & \multicolumn{1}{c|}{\textbf{User}} & \multicolumn{4}{c|}{\textbf{Access \& Interaction}} \\ \cline{2-6}\cline{9-12} 
 & \multirow{2}{*}{\rotatebox[origin=c]{-90}{\faTag}}  & \multirow{2}{*}{\faShield*}  & \multirow{2}{*}{\faClipboardList} & \multirow{2}{*}{\faUnlock} & \multirow{2}{*}{\rotatebox[origin=c]{180}{\faQuestion}}  & \multicolumn{1}{c|}{\textbf{Collection}} & \multicolumn{1}{c|}{\textbf{Tracking \&}} & \multicolumn{1}{c|}{\textbf{App}} & \multicolumn{1}{c|}{\textbf{SDK}} & \multicolumn{1}{c|}{\textbf{App Store}} & \multicolumn{1}{c|}{\textbf{End}} \\
 &  &  &  &  &  & \multicolumn{1}{c|}{\textbf{\& Handling}} & \multicolumn{1}{c|}{\textbf{Fingerprinting}} & \multicolumn{1}{c|}{\textbf{Developer}} & \multicolumn{1}{c|}{\textbf{Developer}} & \multicolumn{1}{c|}{\textbf{Reviewer}} & \multicolumn{1}{c|}{\textbf{User}} \\
\midrule
\rowcolor[HTML]{EFEFEF}
\multicolumn{1}{@{}l}{\rotatebox[origin=c]{-90}{\faTag}~Privacy Label} & \multicolumn{1}{c}{\faList*~\faSun} & \multicolumn{1}{c}{\faList*[regular]~\faSun[regular]} & \multicolumn{1}{c}{} & \multicolumn{1}{c}{} & \multicolumn{1}{c}{} & \multicolumn{1}{c}{\faLink~\faLightbulb~\faLayerGroup} & \multicolumn{1}{c}{} & \multicolumn{1}{c}{\faPenNib} & \multicolumn{1}{c}{} & \multicolumn{1}{c}{\faEye} & \multicolumn{1}{c}{\faEye} \\
\multicolumn{1}{@{}l}{\faShield*~Privacy Manifest} & \multicolumn{1}{c}{\faList*~\faSun} & \multicolumn{1}{c}{\faList*~\faSun} & \multicolumn{1}{c}{\faSun[regular]} & \multicolumn{1}{c}{\faSun[regular]} & \multicolumn{1}{c}{\faList*[regular]~\faSun[regular]} & \multicolumn{1}{c}{\faLink~\faLightbulb~\faLayerGroup} & \multicolumn{1}{c}{\faExclamationTriangle} & \multicolumn{1}{c}{\faPenNib} & \multicolumn{1}{c}{\faPenNib} & \multicolumn{1}{c}{\faCheckCircle} & \multicolumn{1}{c}{} \\
\rowcolor[HTML]{EFEFEF}
\multicolumn{1}{@{}l}{\faClipboardList~Entitlements} & \multicolumn{1}{c}{\faSun[regular]} & \multicolumn{1}{c}{\faSun[regular]} & \multicolumn{1}{c}{\faList*~\faSun} & \multicolumn{1}{c}{\faList*[regular]~\faSun[regular]} & \multicolumn{1}{c}{\faList*[regular]~\faSun[regular]} & \multicolumn{1}{c}{\faLayerGroup} & \multicolumn{1}{c}{\faExclamationTriangle} & \multicolumn{1}{c}{\faPenNib} & \multicolumn{1}{c}{} & \multicolumn{1}{c}{\faCheckCircle} & \multicolumn{1}{c}{} \\
\multicolumn{1}{@{}l}{\faUnlock~Permissions} & \multicolumn{1}{c}{\faSun[regular]} & \multicolumn{1}{c}{\faSun[regular]} & \multicolumn{1}{c}{\faList*[regular]~\faSun[regular]} & \multicolumn{1}{c}{\faList*~\faSun} & \multicolumn{1}{c}{\faList*[regular]~\faSun[regular]} & \multicolumn{1}{c}{\faLayerGroup} & \multicolumn{1}{c}{\faExclamationTriangle} & \multicolumn{1}{c}{\faPenNib} & \multicolumn{1}{c}{} & \multicolumn{1}{c}{} & \multicolumn{1}{c}{\faCheckCircle}  \\
\rowcolor[HTML]{EFEFEF}
\multicolumn{1}{@{}l}{\rotatebox[origin=c]{180}{\faQuestion}~Native API Calls} & \multicolumn{1}{c}{\faSun[regular]} & \multicolumn{1}{c}{\faSun[regular]} & \multicolumn{1}{c}{\faList*[regular]~\faSun} & \multicolumn{1}{c}{\faList*[regular]~\faSun} & \multicolumn{1}{c}{\faList*~\faSun} & \multicolumn{1}{c}{\faLayerGroup} & \multicolumn{1}{c}{\faExclamationTriangle~\faFingerprint} & \multicolumn{1}{c}{\faPenNib} & \multicolumn{1}{c}{\faPenNib} & \multicolumn{1}{c}{} & \multicolumn{1}{c}{}  \\
\bottomrule
\end{tabular}
}
\end{table*}

Table~\ref{tab:threat_model_table} provides a taxonomy of \mac privacy components and disclosure mechanisms, evaluated across four dimensions: (i) overlap in taxonomy and functionality, (ii) data collection and handling practices, (iii) tracking and fingerprinting surfaces, and (iv) access and interaction by different entities (e.g., developers, reviewers, users).

\noindent\textbf{Privacy Labels} are user-facing disclosures that report data collection categories and stated purposes. They rely on the same taxonomy as manifests and permissions, since developers generate them based on bundle-defined components. During the app’s lifecycle, the App Store verifies their presence through internal checks, without however, verifying their accuracy.
As our analysis demonstrated, labels remain developer-declared rather than system-enforced or validated. 
Even if they demonstrate the intended practices, they provide no information about fingerprinting vectors or runtime behavior.

\noindent\textbf{Privacy Manifests} are developer-supplied metadata files embedded in macOS app bundles and SDKs. They adopt identical data-type taxonomy as \nl  and require structured declarations of collected data, stated purposes, and externally accessed tracking domains (e.g., \texttt{NSPrivacyTrackingDomains}). 
During app's submission, Apple’s review pipeline processes  manifests to generate the corresponding \nls. 
However, manifests are solely developer-declarative, since they capture developer-stated practices. 
Similar to \nls they do not specify the application's tracking purposes or runtime functionality.

\noindent\textbf{Entitlements} are the primary access control mechanism on \mac. 
They are embedded in the app bundle’s \texttt{.entitlements} file and define the system services and resources an application can access (e.g., push notifications). 
Entitlements are enforced by macOS at runtime, controlling the app’s effective permissions during execution.

\noindent\textbf{Permissions} are enforced by macOS at runtime when an application requests access to protected resources such as the camera, microphone, or location. Permissions provide the interface through which users must grant or deny access to sensitive components. They are enforced at runtime and directly tied to application behavior once access is requested. By design, they do not provide information about how resources are used once access has been granted.

\noindent\textbf{Native API Calls} are used by applications to deliver their functionality.
They are widely accessible by both the application developer, and also the third-party SDKs.
The APIs invoked during runtime not only reflect the data uses declared by the application, and also expose additional invocations that may be exploited for tracking or other invasive practices.
This runtime information can only be obtained through dynamic analysis, which requires additional effort for execution monitoring and attribution.

Overall, each mechanism contributes a complementary view of application behavior, but none provides an accurate representation of privacy practices.
Moreover, there is no centralized validation mechanism (e.g., the Mac App Store) to enforce consistency.
Our analysis demonstrates the fundamental limitations of the current validation process, while our system performed \emph{end-to-end} analysis to detect inconsistencies in disclosure practices.

\begin{figure}[t!]
\centering
\centering
\begin{lstlisting}[language=Swift,firstnumber=1,basicstyle=\footnotesize, caption=Example Swift code to capture a video on \mac., float=false, label={lst:example-camera-access}]
var isAuthorized: Bool {
  get async {
    let status = AVCaptureDevice.authorizationStatus(for: .video)
    // Determine if the user previously authorized camera access.
    var isAuthorized = status == .authorized
    // If the system hasn't determined the user's authorization status,
    // explicitly prompt them for approval.
    if status == .notDetermined {
        isAuthorized = await AVCaptureDevice.requestAccess(for: .video)
    } return isAuthorized}}
func setUpCaptureSession() async {
  // Check for camera authorization.
  guard await isAuthorized else { return }
  // Create the capture session.
  let captureSession = AVCaptureSession()
  // Find the default video device.
  guard let videoDevice = AVCaptureDevice.default(for: .video) else { return }
  do {
    // Wrap the video device in a capture device input.
    let videoInput = try AVCaptureDeviceInput(device: videoDevice)
    // If the input can be added, add it to the session.
    if captureSession.canAddInput(videoInput) {
        captureSession.addInput(videoInput)
    }} catch {
    // Configuration failed. Handle error. }}
\end{lstlisting}
\end{figure}

\begin{figure}[t]
\begin{lstlisting}[language=json,firstnumber=1,basicstyle=\footnotesize, caption=Example unified declaration for camera access., label={lst:unified-declaration}]
"Camera": {
  "Data Type": {"Video": true},
  "Purpose": {"App Functionality": true},
  "Handling": {"Anonymization": false},
  "User Permission": {
   "Required": true, "Status": .notdetermined},
  "API Access": [
    "AVCaptureSession",
    "AVCaptureDevice",
    "AVCaptureVideoDataOutput"]}
\end{lstlisting}
\end{figure}

%
%
%

\section{\system : Nutrition Label to Entitlement mapping}
\label{appendix:nl_enti}

\begin{table*}
\centering
\caption{Nutrition Label to Entitlement and API mapping}
\label{tab:exfil_app2}
\resizebox{1.0\textwidth}{!}{%
\begin{tabular}{l|p{0.45\textwidth}|p{0.55\textwidth}}
\toprule
\textbf{Label} & \textbf{Entitlement} & \textbf{API} \\
\midrule

\rowcolor{lightgray}
user id &
accounts.appleaccount.fullaccess, developer.game-center, developer.ubiquity-kvstore-identifier, developer.upi-device-validation &
gethostname, hostinfo, NSUserName, NSHomeDirectory, NSFileManager, NSHost, IOBluetoothDevice, NSAppleScript, NSTask, UIDevice \\
\midrule

email address &
accounts.appleaccount.fullaccess, developer.mail-client, mail &
ABCopyArrayOfAllPeople, ABGetSharedAddressBook, ABAddressBook, CNContactSaveRequest, CNContactStore, NSAppleScript, CNContactFetchRequest \\
\midrule

\rowcolor{lightgray}
phone number &
accounts.appleaccount.fullaccess, private.contactsui &
ABCopyArrayOfAllPeople, ABGetSharedAddressBook, ABAddressBook, CNContactSaveRequest, CNContactStore, EKEventStore, NSAppleScript, CNContactFetchRequest \\
\midrule

\multirow{2}{*}{name} &
accounts.appleaccount.fullaccess, private.contactsui, security.personal-information.addressbook, security.temporary-exception.files.home-relative-path.read-only &
ABCopyArrayOfAllPeople, ABGetSharedAddressBook, NSUserName, NSHomeDirectory, NSHost, \\
& & ABAddressBook, CNContactSaveRequest, CNContactStore, EKEventStore, NSAppleScript, CNContactFetchRequest \\
\midrule

\rowcolor{lightgray}
physical address &
accounts.appleaccount.fullaccess, private.contactsui, security.personal-information.addressbook &
ABCopyArrayOfAllPeople, ABGetSharedAddressBook, ABAddressBook, CNContactSaveRequest, CNContactStore, EKEventStore, NSAppleScript, PKPaymentRequest, CNContactFetchRequest \\
\midrule

other user content &
developer.contacts.notes, developer.ubiquity-container-identifiers, developer.ubiquity-kvstore-identifier, developer.user-management, security.files.downloads.read-only, security.files.downloads.read-write, security.files.user-selected.read-only, security.files.user-selected.read-write, security.personal-information.calendars &
LSSharedFileListCopySnapshot, EKEventStore, NSAppleScript \\
\midrule

\rowcolor{lightgray}
browsing history &
developer.web-browser &
LSSharedFileListCopySnapshot, NSAppleScript \\
\midrule

device id &
accounts.appleaccount.fullaccess, developer.device-information.user-assigned-device-name, security.temporary-exception.files.home-relative-path.read-only &
AudioDeviceID, DCDevice, AVCaptureDevice, CWInterface, NSAppleScript, NSTask, UIDevice \\
\midrule

\rowcolor{lightgray}
precise location &
developer.location.push, security.personal-information.location &
CLLocationManager, LocationManagerDelegate, NSAppleScript \\
\midrule

coarse location &
developer.location.push, developer.networking.networkextension, security.personal-information.location &
CLLocationManager, LocationManagerDelegate, NSUserDefaults, NSTimeZone, NSAppleScript \\
\midrule

\rowcolor{lightgray}
payment info &
accounts.appleaccount.fullaccess, developer.financekit, developer.in-app-payments &
PKPaymentRecurringRequest, PKPayment, SKMutablePayment, NSAppleScript, PKPaymentRequest \\
\midrule

credit info &
\multirow{3}{*}{accounts.appleaccount.fullaccess} &
PKPaymentRecurringRequest, PKPayment, SKMutablePayment, NSAppleScript, PKPaymentRequest \\
\midrule

\rowcolor{lightgray}
emails or text messages &
accounts.appleaccount.fullaccess, developer.mail-client, mail &
MEMessage, MSConversation, MEMessageState, NSAppleScript \\
\midrule

performance data & 
& 
MXCellularConditionMetric, MXCPUMetrics, MXGPUMetrics, NSAppleScript, MXAppLaunchDiagnostic, MXAppExitMetric, MXNetworkTransferMetric, MXCPUExceptionDiagnostic, MXDiskIOMetric, MXMemoryMetric, MXDisplayMetric \\
\midrule

\rowcolor{lightgray}
advertising data & 
& 
ASIdentifierManager, NSAppleScript, SKAdNetwork \\
\midrule

photos or videos &
security.assets.pictures.read-only, security.assets.pictures.read-write, security.device.camera, security.personal-information.photos-library &
AVCaptureSession, NSAppleScript, PHFetchResultChangeDetails, PHFetchResult \\
\midrule

\rowcolor{lightgray}
audio data &
security.device.audio-input, security.device.microphone &
AVCaptureSession, NSAppleScript, PHFetchResultChangeDetails, PHFetchResult \\
\midrule

purchase history &
developer.in-app-payments &
NSAppleScript \\
\midrule

\rowcolor{lightgray}
product interaction & 
& 
NSAppleScript \\
\midrule

crash data & 
& 
NSAppleScript, MXAppLaunchDiagnostic, MXAppExitMetric, MXCPUExceptionDiagnostic, MXDiskIOMetric, MXMemoryMetric, MXDisplayMetric \\
\midrule

\rowcolor{lightgray}
customer support & 
& 
NSAppleScript \\
\midrule

other financial info &
developer.financekit &
NSAppleScript \\
\midrule

\rowcolor{lightgray}
gameplay content & 
& 
NSAppleScript \\
\midrule

sensitive info &
developer.healthkit.access, developer.healthkit &
NSAppleScript \\
\bottomrule
\end{tabular}
}
\end{table*}

Table~\ref{tab:exfil_app2} maps Nutrition Label categories to the entitlements and APIs that provide access to the corresponding data.
We note that this information \emph{does not} appear in Apple’s documentation and was constructed from official references during our analysis.
As shown in \S\ref{sec:evaluation}, individual label categories correspond to diverse entitlements and APIs.
For example, \texttt{user id} is reported under Apple account and Game Center entitlements and is accessible through host and device APIs.
Multiple categories, including advertising data, performance data, and crash data, have no associated entitlement and can only be accessed through APIs.
This mapping demonstrates that the Nutrition Label taxonomy is technically grounded in concrete entitlements and APIs.
However, the absence of corresponding documentation and guidelines leads to developer confusion, resulting in the \nl violations and discrepancies, as our analysis demonstrated.

\section{Entitlement oriented Discrepancies}
\label{appendix:entitlement_discrepancies}

Table~\ref{tab:extended} expands Table~\ref{tab:entl_discrepancies} (\S\ref{subsec:bundle_analysis}) with the entitlement-to-\nl mapping and the Nutrition Label data types they correspond to.
As our analysis highlights, multiple entitlements map to different and diverse \nl categories.
For example, \texttt{appleaccount.fullaccess} maps to user and device identifiers, contact details, communications, and payment information.
The largest discrepancies occur for Identifiers, Name and Contact Information, Photos or Videos, Health, and Financial Information, which are frequently omitted even though the entitlements indicate access.
Overall, this result demonstrates that the impact of discrepancies is greater when entitlements  map to multiple \nl categories are omited.

\begin{table*}[h!]
\centering
\caption{Distribution of apps with discrepancies between declared entitlements and corresponding Nutrition Label data types.}
\label{tab:extended}
  \begin{tabular}{p{5cm}cccp{5cm}}
    \toprule
    \textbf{Entitlement} & \textbf{\# TotalApp (\%)} &  \textbf{\# TopApp (\%)} & \textbf{\# Rand.App (\%)} &  \textbf{Label Data Type} \\
    \midrule
    \multicolumn{5}{c}{\textbf{com.apple.security}} \\
    \hline
    \rowcolor{lightgray} files.user-selected.read-only  & 95\% & 94.1\% & 95.7\%  & Other User Content  \\
    personal-information.location  & 93.6\% & 93.8\% & 93.4\% & Precise Location, Coarse Location  \\
    \rowcolor{lightgray} personal-information.addressbook  & 93.8\% & 93.9\% & 92.8\%  & Name, Physical Address  \\
    files.downloads.read-write  & 92.3\% & 93.8\% & 89.6\%  & Other User Content  \\
    \rowcolor{lightgray} personal-information.photos-library  & 90.8\% & 90\% & 92.1\%  & Photos or Videos  \\
    files.downloads.read-only  & 88.8\% & 66.6\% & 100\% & Other User Content  \\
    \rowcolor{lightgray} files.user-selected.read-write  & 90.7\% & 87.7\% & 94.5\% & Other User Content \\
    device.microphone  & 89.4\% & 88.2\% & 90.4\% & Audio Data  \\
    \rowcolor{lightgray} assets.pictures.read-write  & 87.1\% & 89.4\% & 83.3\%  & Photos or Videos  \\
    personal-information.calendars  & 86.6\% & 84.7\% & 92.3\% & Other User Content  \\
    \rowcolor{lightgray} device.audio-input  & 86.1\% & 87.6\% & 84.5\%  & Audio Data  \\
    device.camera  & 84.5\% & 84.5\% & 84.4\%  & Photos or Videos \\
    \rowcolor{lightgray} assets.pictures.read-only  & 80.9\% & 75\% & 100\% & Photos or Videos \\
    temporary-exception.files.home-relative-path.read-only  & 68.7\% & 61.5\% & 100\%  & Device ID, Name  \\
    \midrule
    \multicolumn{5}{c}{\textbf{com.apple.developer}} \\
    \hline
    healthkit  & 100\% & 100\% & 100\%  & Health, Sensitive Info \\
     \rowcolor{lightgray} healthkit.access  & 100\% & 100\% & 100\%  & Health, Sensitive Info \\

    ubiquity-kvstore-identifier  & 93.6\% & 90.5\% & 100\%  & Other User Content \\
     \rowcolor{lightgray} networking.networkextension  & 90.3\% & 89.4\% & 91.6\% & Coarse Location  \\
    ubiquity-container-identifiers  & 90.7\% & 87.1\% & 97.3\%  & Other User Content  \\
     \rowcolor{lightgray} in-app-payments  & 66.6\% & 50\% & 100\%  & Payment Info, Purchase History  \\
    game-center  & 65\% & 66.6\% & 64\%  & User ID \\
     \rowcolor{lightgray} financekit  & 50\% & 50\% & N.A.  & Payment Info, Other Financial Info \\
    \midrule
    \multicolumn{5}{c}{\textbf{com.apple}} \\
    \hline
    \rowcolor{lightgray} mail  & 100\% & 100\% & 100\%  & Email Address, Emails or Text Messages  \\
    accounts.appleaccount.fullaccess  & 100\% & N.A. & 100\%  & User ID, Device ID, Phone Number, Name, Physical Address, Emails or Text Messages, Payment Info, Credit Info, Health, Fitness \\
    \bottomrule
\end{tabular}
\end{table*}

\section{Unified Privacy Declaration}
\label{appendix:unified}

In Listing~\ref{lst:example-camera-access} we include a snippet of example
Swift code that a developer may use to capture enable a user to capture video
using the device camera. The code checks for camera access authorization,
requests access if not determined, and sets up a capture session if authorized,
before creating a recording session.

Listing~\ref{lst:unified-declaration} demonstrates how  how multiple fragmented
disclosures, that are currently spread across entitlements, privacy manifests,
and user-facing labels, can be merged into a single, unified declaration.  A
developer, may specify the data type collected (video), it's purpose
(functionality), the handling practice (no anonymization) and other metadata
about how APIs are invoked.  This unified structure allows the same declaration
to serve simultaneously for OS-level enforcement, App Store review, and user
transparency. 

%
%
%

\end{document}